\documentclass[aps,prd,floatfix,nofootinbib,superscriptaddress,preprint]{revtex4-1}
\usepackage{amssymb}
\usepackage{amsmath}
\usepackage{amsfonts}
\usepackage{graphicx}
\usepackage{color}
\usepackage{xspace}
\usepackage{ulem}
\usepackage{amsmath,amssymb,amsfonts,latexsym,cancel}
\usepackage{mathrsfs}
\usepackage{xcolor}                 
\usepackage{cancel}
\usepackage{anysize}
\usepackage{wrapfig}
\usepackage{slashed}
\usepackage{lipsum}
\usepackage{multirow}
\usepackage{physics}
\usepackage{setspace}
\usepackage{ulem}
\usepackage{stackengine}
\usepackage{xr}

\newcommand{\AddrUCL}{
Department of Physics and Astronomy, University College London,\\
London WC1E 6BT, United Kingdom
}

\newcommand{\AddrSISSA}{
SISSA, International School for Advanced Studies, \\
INFN, Sezione di Trieste, Via Bonomea 265, I-34136 Trieste, Italy
}

\newcommand{\AddrUG}{
Institute of Physics, NAWI Graz, University of Graz, \\
Universit\"atsplatz 5, A-8010 Graz, Austria
}

\newcommand{\AddrTUM}{
 Physik Department T70, Technische Universit\"at M\"unchen,\\
James-Franck-Stra{\ss}e 1, D-85748 Garching, Germany
}

\begin{document}

\title{Probing Active-Sterile Neutrino Transition Magnetic Moments with Photon Emission from CE$\nu$NS} 

\author{Patrick D. Bolton} 
\email{patrick.bolton@ts.infn.it}\affiliation{\AddrUCL}\affiliation{\AddrSISSA}
\author{Frank F. Deppisch} 
\email{f.deppisch@ucl.ac.uk}\affiliation{\AddrUCL}
\author{K\aa re Fridell} 
\email{kare.fridell@tum.de}\affiliation{\AddrTUM}
\author{Julia Harz} 
\email{julia.harz@tum.de}\affiliation{\AddrTUM}
\author{Chandan Hati} 
\email{c.hati@tum.de}\affiliation{\AddrTUM}
\author{Suchita Kulkarni} 
\email{suchita.kulkarni@uni-graz.at}\affiliation{\AddrUG}

\begin{abstract}
\noindent In the presence of transition magnetic moments between active and sterile neutrinos, the search for a Primakoff upscattering process at coherent elastic neutrino-nucleus scattering (CE$\nu$NS) experiments can provide stringent constraints on the neutrino magnetic moment. We show that a radiative upscattering process with an emitted photon in the final state can induce a novel coincidence signal at CE$\nu$NS experiments that can also probe neutrino transition magnetic moments beyond existing limits. Furthermore, the differential distributions for such a radiative mode can also potentially be sensitive to the Dirac vs. Majorana nature of the sterile state mediating the process. This can provide valuable insights into the nature and mass generation mechanism of the light active neutrinos.
\end{abstract}

\maketitle

\section{Introduction}
\label{sec:intro}

Coherent elastic neutrino-nucleus scattering (CE$\nu$NS)~\cite{Freedman:1973yd} was first observed in 2017 by the COHERENT collaboration with a statistical significance of 6.7$\sigma$~\cite{COHERENT:2017ipa}. Future CE$\nu$NS experiments with extremely low nuclear recoil thresholds now aim to detect neutrino-nucleus scattering events with $\mathcal{O}$(eV) momentum transfers~\cite{NUCLEUS:2019kxv, NUCLEUS:2019igx}. These next-generation experiments will not only test the Standard Model (SM) CE$\nu$NS rate more precisely but also constrain physics beyond the SM. One such new physics (NP) scenario is the so-called Primakoff upscattering of light active neutrinos to heavy sterile neutrinos via a transition magnetic moment. This \textit{dipole portal} could be a promising way to probe both the existence of sterile neutrinos and possible NP generating the dipole coupling. 

The focus of the present work is a closely-related process that can also produce distinct signatures at CE$\nu$NS experiments. This is the upscattering of an incoming active neutrino to a sterile neutrino, which subsequently decays to an active neutrino and photon, see Fig.~\ref{fig:feyndiag}. Searches for such a \textit{radiative} upscattering process have been suggested for the DUNE~\cite{Schwetz:2020xra}, IceCube~\cite{Coloma:2017ppo}, and Super-Kamiokande~\cite{Atkinson:2021rnp} experiments. This process can be used to probe transition magnetic moments in CE$\nu$NS experiments as well as to distinguish sterile neutrinos of Dirac or Majorana nature, indicating the corresponding nature of the active neutrinos~\cite{Voloshin:1987qy, Pal:1981rm, Shrock:1982sc, Bell:2005kz, Cepedello:2018zvr, Magill:2018jla, Bolton:2019bou, Brdar:2020quo, Babu:2020ivd}. 

\section{Neutrino Transition Magnetic Moments}
\label{sec:eft}

A transition magnetic moment between the three active neutrinos and a sterile neutrino (or gauge-singlet fermion) can be described by the effective Lagrangian,
\begin{align}
\label{eq:Ld}
\mathcal{L} \supset \frac{\mu^\alpha_{\nu N}}{2}\bar{\nu}_{\alpha L} \sigma_{\mu\nu} P_R N F^{\mu\nu} + \frac{\mu_{N' N}}{2} \bar{N'}\sigma_{\mu\nu}P_R N F^{\mu\nu} +\text{h.c.}\,,
\end{align}
where $F^{\mu \nu} = \partial^\mu A^\nu - \partial^\nu A^\mu$ is the electromagnetic field strength tensor, $\nu_{\alpha L}$ is an active neutrino field of flavour $\alpha = \{e,\mu,\tau\}$, $N$ is the sterile neutrino field, and $\mu^\alpha_{\nu N}$ are the active-sterile dipole couplings. Here, we introduce $N'$ as an additional light ($m_{N'}\ll m_N$) sterile state with a sterile-sterile transition magnetic moment to $N$ (with a transition dipole coupling $\mu_{N'N}$). The Lagrangian in Eq.~\eqref{eq:Ld} is only valid at energies below the electroweak (EW) scale because it is not invariant under the SM gauge symmetry. Above the EW scale, it must be matched onto operators at dimension-six and above containing the $\text{SU}(2)_L$ and $\text{U}(1)_Y$ gauge fields, SM Higgs doublet $H$, and lepton doublet $L_\alpha$. Coherent scattering processes take place at energies well below the EW scale and hence Eq.~\eqref{eq:Ld} remains applicable.

The active and sterile neutrinos in Eq.~\eqref{eq:Ld} can either be Dirac or Majorana fermions. For example, the active neutrinos $\nu_{\alpha L}$ can either be the left-handed Weyl components of Dirac ($\nu_\alpha = \nu_{\alpha L} + \nu_{\alpha R}$) or Majorana ($\nu_\alpha = \nu_{\alpha L} + \nu^c_{\alpha L}$) fields, where in the former case it is necessary to introduce additional (sterile) Weyl fields $\nu_{\alpha R}$. Similarly, the sterile neutrino $N$ (and $N'$) can either be composed of independent left- and right-handed Weyl fields ($N = N_L + N_R$), or a single right-handed field ($N = N^c_R + N_R$). In principle, there are four possible combinations of Dirac and Majorana active and sterile neutrinos. However, it can be shown that for energy scales much greater than the active neutrino masses, $E_\nu \gg m_\nu$, the rates for processes involving Dirac and Majorana active neutrinos are approximately equal, in accordance with the \textit{Dirac-Majorana confusion theorem}~\cite{Kayser:1981nw, Kayser:1982br}. In this work, we will consider sterile neutrinos with masses similar to the energy scale of the process, $E_N \sim m_N$. Consequently, the difference between the rates for Dirac or Majorana sterile neutrinos can be significant.

The active-sterile dipole couplings $\mu^\alpha_{\nu N}$ in Eq.~\eqref{eq:Ld} give rise to the Primakoff upscattering process $\nu_\alpha A \to N A$; see, e.g., Refs.~\cite{Vogel:1989iv,Balantekin:2013sda,Magill:2018jla}. For relativistic $\nu_\alpha$, it can be shown that the differential cross section in the nuclear recoil energy $E_R$ for this process is the same for outgoing Dirac and Majorana $N$. The non-observation of deviations from the SM CE$\nu$NS nuclear recoil rate at experiments can therefore be used to constrain the dipole couplings $\mu^\alpha_{\nu N}$ as a function of the sterile neutrino mass $m_N$. Constraints on the dipole couplings $\mu^\alpha_{\nu N}$ have been set by a variety of experiments (c.f. Fig.~\ref{fig:nudp_constraints}) and are generally flavour dependent~\cite{Schwetz:2020xra}.

\begin{figure}[t!]
	\centering
	\includegraphics[width=7cm]{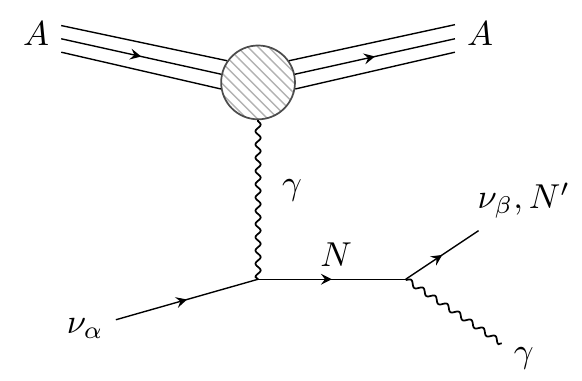}
\caption{Radiative upscattering process $\nu_\alpha A\to \nu_\beta A\gamma$ ($\nu_\alpha A\to N' A\gamma$) via the dipole couplings $\mu^\alpha_{\nu N}$ and $\mu^\beta_{\nu N}$ ($\mu_{N' N}$) and an intermediate Dirac or Majorana sterile neutrino $N$.}
\label{fig:feyndiag}
\end{figure}

The active-sterile dipole couplings $\mu^\alpha_{\nu N}$ can also induce the \textit{radiative} upscattering process $\nu_\alpha A \to \nu_\beta A \gamma$, depicted in Fig.~\ref{fig:feyndiag}. The rate of the process is proportional to $|\mu^\alpha_{\nu N}\mu^\beta_{\nu N}|^2$ and is therefore suppressed with respect to the Primakoff upscattering. However, the outgoing photon serves as an additional signal to the nuclear recoil and provides kinematical information that can be used to discriminate between Dirac and Majorana $N$. We would like to remark that because the outgoing neutrino is not detected, the actual process is $\nu_\alpha A \to X A \gamma$, where $X$ may be an active neutrino $\nu_\beta$ or a sterile state $N'$. The rate is then generically proportional to $|\mu^\alpha_{\nu N} \sum_X \mu_{X N}|^2$, where the sum is over all active and sterile neutrinos that are coupled to $N$ via a transition magnetic moment.

In this work we will examine as a case study the upcoming NUCLEUS experiment located at the Chooz reactor facility~\cite{NUCLEUS:2019kxv, NUCLEUS:2019igx}. The planned experimental arrangement and future upgrade make it feasible to detect the energies and angles of outgoing photons~\cite{NUCLEUS}. The (non)-detection of photons at NUCLEUS can thus constrain the dipole couplings $\mu^\alpha_{\nu N}$ and (if photons are detected) provide important information towards identifying the nature and mass generation mechanism of the active neutrinos.

\section{Energy Distributions as a Novel Probe for Dirac vs. Majorana Neutrinos}
\label{sec:calculation}

We give the detailed calculation of the differential cross section for $\nu_\alpha A\to X A\gamma$ in Appendix~\ref{sec:app_calc}. Here, we outline the main results and the salient features relevant to our study. The amplitudes for the $\nu_\alpha A\to X A\gamma$ process are given in the scenarios where the sterile neutrino $N$ is Dirac or Majorana, respectively, by
\begin{align}
\label{eq:M_dirac}
&i\mathcal{M}^{\text{D}}_{\nu_\alpha A\to X A \gamma} = \mu_{\nu N}^\alpha \mu_{X N} [\bar{u}_{X}\sigma_{\mu\nu} P_R (\slashed{p}\hspace{-0.5em}\phantom{p}_N+m_N)\sigma_{\rho\sigma}P_L u_{\nu_\alpha}] F^{\mu\nu\rho\sigma}\,, \\
&i\mathcal{M}^{\text{M}}_{\nu_\alpha A\to X A \gamma} = \mu_{\nu N}^\alpha \mu_{X N} [\bar{u}_{X}\sigma_{\mu\nu} (\slashed{p}\hspace{-0.5em}\phantom{p}_N+m_N)\sigma_{\rho\sigma}P_L u_{\nu_\alpha}] F^{\mu\nu\rho\sigma}\,,
\label{eq:M_maj}
\end{align}
where $F^{\mu\nu\rho\sigma}$, given in Appendix~\ref{sec:app_calc}, contains the denominator of the sterile neutrino propagator, the hadronic current of the nucleus $A$, the propagator of the exchanged photon, and the four-momentum and polarisation of the outgoing photon. We have taken the dipole couplings to be imaginary, $(\mu_{X N})^* = -\mu_{X N}$, and thus CP conserving if $X$ and $N$ have opposite CP phases~\cite{Giunti:2014ixa}. In the Dirac case, $N$ is created by the hermitian conjugate of the first term and annihilated by the second term in Eq.~\eqref{eq:Ld}. In the Majorana case, both the second term in Eq.~\eqref{eq:Ld} and its hermitian conjugate annihilate $N$; it is then possible to make the replacement $P_R \to (P_R+P_L) = 1$ at the decay vertex. The $P_R$ and $P_L$ project out the momentum $\slashed{p}\hspace{-0.5em}\phantom{p}_N$ and mass $m_N$ terms from the $N$ propagator, respectively. The three-body phase space of the final state $X A\gamma$ is described by four variables; we choose the nuclear recoil energy $E_R$, photon energy $E_\gamma$, nuclear recoil angle $\theta_R$, and photon angle $\theta_\gamma$ (both angles defined with respect to the incoming neutrino direction). 

For a CE$\nu$NS experiment to detect an outgoing photon, and therefore the process $\nu_\alpha A\to X A\gamma$, a radiative sterile neutrino decay $N\to X\gamma$ must take place within the detector. The probability for this to occur is given by the $N\to X\gamma$ branching ratio $\mathcal{B}_{N\to X\gamma}=\Gamma_{N\to X\gamma}/\Gamma_N$ multiplied by the probability for a decay to take place within the detector, $P_{N}^{\text{det}}=1-\text{exp}(-\frac{L_{\text{det}} \Gamma_{N}}{\beta\gamma})$, where $L_{\text{det}}$ is the detector length, $\Gamma_N$ is the total decay width of $N$, and the boost factors are given by $\beta\gamma = \sqrt{\gamma^2-1}$ and $\gamma = E_N/m_N$. It is also possible that $N$ decays via invisible channels (for example, to a light dark sector), which contribute to the total width as $\Gamma_N = \Gamma_{N\to X\gamma} + \Gamma_N^{\text{inv}}$, giving
\begin{align}
\mathcal{B}_{N\to X\gamma}P_{N}^{\text{det}}=\frac{\Gamma_{N\to X\gamma}}{\Gamma_N}\bigg[1-\text{exp}\bigg(-\frac{L_{\text{det}} \Gamma_{N}}{\beta\gamma}\bigg)\bigg]
\end{align}
as the probability of observing a radiative decay inside the detector. For example, for $m_{N}\sim 1~\mathrm{MeV}$, the upper limit on the active-sterile mixing squared from beta decays is $|V_{e N}|^2\lesssim 10^{-2}$, implying a rate of $\Gamma_{N\to 3\nu} = \frac{G_{F}^2m_N^5}{96\pi^3}|V_{eN}|^2 \lesssim 10^{-27}~\mathrm{MeV}$ for the invisible decay of $N$ to three light neutrinos. The Primakoff upscattering process is observed, through the nucleon recoil, if there is an invisible decay, or a radiative decay occurs outside the detector.

Assuming $\Gamma_N = \Gamma_{N\to X\gamma} + \Gamma_N^{\text{inv}}$ to be small and therefore the total $N$ decay length to be much longer than the detector length, $\ell_N = \beta\gamma \tau_N = \beta\gamma/\Gamma_N\gg L_{\text{det}}$, the exponential in the decay probability $P_{N}^{\text{det}}$ can be Taylor expanded to give $\mathcal{B}_{N\to X\gamma}P_{N}^{\text{det}} = \Gamma_{N\to X\gamma}L_{\text{det}}/(\beta\gamma)$, which is independent of the total decay width. For a detector length of $L_{\text{det}} = 5$~cm, the rate of $\nu_\alpha A\to X A\gamma$ events is independent of an invisible $N$ decay width up to $\Gamma^{\text{inv}}_N \sim 10^{-11}~\text{MeV} $. Above this value, $P_{N}^{\text{det}}  \approx 1$ and the probability that $\nu_\alpha A\to X A\gamma$ occurs within the detector is suppressed by the small branching ratio.

We therefore consider three benchmark scenarios in this work. The first is to assume that $N\to \nu_\beta\gamma$ is the only $N$ decay channel. The branching ratio is $\mathcal{B}_{N\to X\gamma} = 1$ and the decay probability, $P_{N}^{\text{det}} = \Gamma_{N\to X\gamma}L_{\text{det}}/(\beta\gamma)\ll 1$. The second is to introduce an invisible $N$ decay width of size $\Gamma^{\text{inv}}_N = \beta\gamma/L_{\text{det}}\gg \Gamma_{N\to X\gamma}$. The branching ratio is now suppressed, $\mathcal{B}_{N\to X\gamma} \approx \Gamma_{N\to X\gamma}L_{\text{det}}/(\beta\gamma)\ll 1$, while the decay probability is large, $P_{N}^{\text{det}} \approx 0.63$. However, the product $\mathcal{B}_{N\to X\gamma}P_{N}^{\text{det}}$ is roughly the same as in the previous scenario. The third case is to include an additional light sterile state $N'$ ($m_{N'} \ll m_N$) and a non-zero sterile-sterile dipole coupling $\mu_{N'N}$, but no invisible modes. Interestingly, the rate for $N\to N'\gamma$ satisfies $\Gamma_{N\to N'\gamma} \gtrsim \beta\gamma/L_{\text{det}}$  for $\mu_{N' N} \gtrsim 10^{-4}~\mu_B$ and $L_{\text{det}}=5$~cm, while the upper limit on the sterile-sterile transition dipole coupling from invisible vector meson decays ($\phi\to\text{invisible}$) is $\mu_{N' N} \lesssim 5\times 10^{-4}~\mu_B$~\cite{Li:2020lba}. However, stronger bounds of order $\mu_{N' N} \lesssim 10^{-6}~\mu_B$ can be derived from LEP~\cite{Magill:2018jla}. For $\mu_{N' N}\sim 10^{-6}~\mu_B$, the decay probability is much larger than for $\mu^\beta_{\nu N}\sim 10^{-8}~\mu_B$, increasing the expected rate for the radiative signal. We want to emphasise that in this scenario the radiative upscattering $\nu_\alpha A\to X A\gamma$ can compete with Primakoff upscattering $\nu_\alpha A\to N A$ in constraining the active-sterile dipole coupling.

Regardless of the benchmark choice above, when the sterile neutrino total decay width $\Gamma_N$ is much smaller than the sterile neutrino mass $m_{N}$, it is possible to use the narrow width approximation (NWA). In this limit, the $\nu_\alpha A\to XA\gamma$ differential cross section in $E_R$ can be decomposed in the NWA as the $\nu_\alpha A\to N A$ cross section multiplied by the $N\to X\gamma$ branching ratio,
\begin{align}
\label{eq:cross_sec_NW}
\frac{d\sigma^{\text{D}(\text{M})}_{\nu_\alpha A\to X A \gamma}}{dE_R}\bigg|_{\text{NWA}} = \frac{d\sigma_{\nu_\alpha A\to N A}}{dE_R}\frac{\Gamma^{\text{D}(\text{M})}_{N\to X\gamma}}{\Gamma_N}\,,
\end{align}
where $X = \{\nu_\beta,N',\ldots\}$. The differential cross section in $E_R$ therefore has the same shape as for $\nu_\alpha A\to N A$ but is suppressed by an additional factor of $|\mu_{X N}|^2$. To yield a differential rate, Eq.~\eqref{eq:cross_sec_NW} must be multiplied by the incoming neutrino flux $\frac{d\phi_{\nu_\alpha}}{dE_\nu}$ and decay probability $P_{N}^{\text{det}}$ and integrated over the incoming neutrino energy. The decay rate for Majorana $N$ is twice that for Dirac $N$, i.e. $\Gamma^{\text{M}}_{N\to X\gamma} = 2\Gamma^{\text{D}}_{N\to X\gamma} = \frac{m_N^3|\mu_{XN}|^2}{4\pi}$. This is because both decay channels $N\to X\gamma$ and $N\to \bar{X}\gamma$ are open for Majorana $N$, while only the first is open for Dirac $N$.
It follows that the differential cross section in Eq.~\eqref{eq:cross_sec_NW} multiplied by the decay probability is also twice as large for Majorana $N$ compared to Dirac $N$ for $\beta\gamma/\Gamma_N \gg L_{\text{det}}$. However, this difference cannot be used to distinguish the Dirac vs. Majorana nature of $N$, as an overall factor of two can absorbed into the measured value of $\mu_{XN}$.

The $\nu_\alpha A\to X A\gamma$ differential cross sections in the photon energy $E_{\gamma}$ and angle $\theta_{\gamma}$ can also be computed in the NWA, but cannot be factorised as in Eq.~\eqref{eq:cross_sec_NW}. The double differential cross section in these two variables can be written as
\begin{align}
\label{eq:double_diff}
\frac{d^2\sigma^{\text{D}(\text{M})}_{\nu_\alpha A\to X A \gamma}}{dE_\gamma d\theta_\gamma}\bigg|_{\text{NWA}} &= \frac{|\mu^\alpha_{\nu N}\mu_{X N}|^2\alpha Z^2 E_\gamma\sin\theta_\gamma}{128\pi^2 m_A E_\nu m_N \Gamma_N }\int^{t^{+}_1}_{t^{-}_1} dt_1 \, \frac{L_{\mu\nu}^{\gamma,\,\text{D(M)}}H^{\mu\nu} \mathcal{F}^2(t_1)}{t_1^2 \sqrt{-\Delta_4}}\bigg|_{s_1 = m_N^2}\,,
\end{align}
where $L_{\mu\nu}^{\gamma,\text{D(M)}}H^{\mu\nu}$ is the contraction of leptonic and hadronic currents for the process and $\Delta_4$ is the $4\times4$ symmetric Gram determinant formed from any four of the five incoming or outgoing four-momenta, both defined in Appendix~\ref{sec:app_calc}. Here, the variable $t_1 = q^2 = -2 m_A E_R$ corresponds to the four-momentum squared of the exchanged photon, $s_1 = p_N^2$ to the four-momentum squared of the sterile neutrino, and $\mathcal{F}(t_1)$ is the nuclear form factor. The limits of integration $t_1^{\pm}$ are found by solving $\Delta_4 = 0$ for $t_1$. In general, the integral cannot be performed analytically due to the presence of $\mathcal{F}(t_1)$.

\begin{figure*}[t!]
    \centering
    \includegraphics[width=0.9\textwidth]{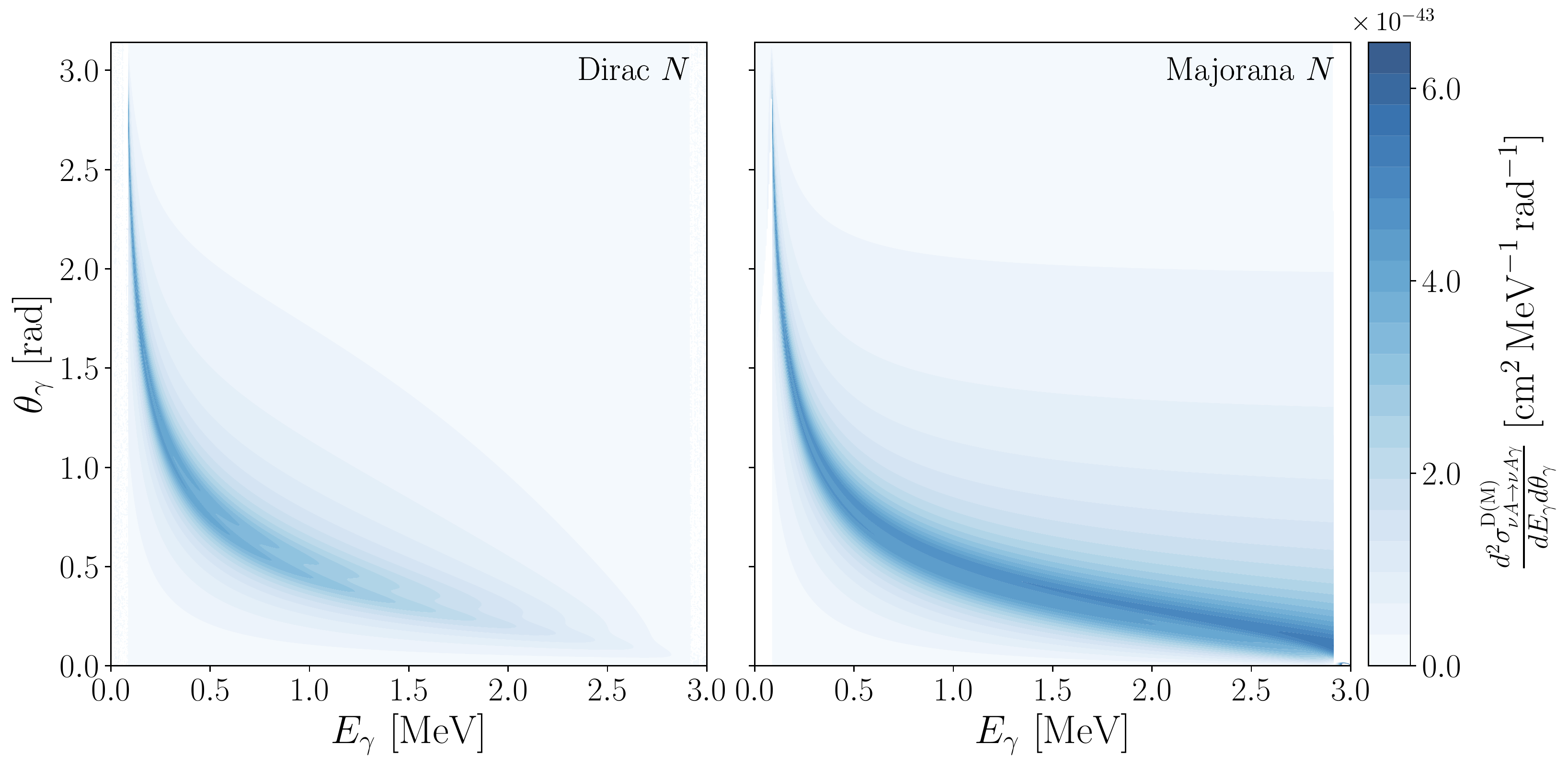}
    \caption{Double differential cross sections for the process $\nu_\alpha A\to \nu_\alpha A \gamma$ in the outgoing photon energy $E_{\gamma}$ and angle $\theta_{\gamma}$ for an incoming neutrino of energy $E_\nu = 3\,\,\mathrm{MeV}$ scattering from a $^{73}$Ge nucleus and an intermediate sterile neutrino of mass $m_{N} = 1\,\,\mathrm{MeV}$. We choose the values $\mu^\alpha_{\nu N} = 3\times 10^{-8}$~$\mu_B$ and $\Gamma_N = 10^{-11}$~MeV. The cases for Dirac and Majorana $N$ are shown to the left and right, respectively.}
    \label{fig:contour_plots}
\end{figure*}

In Fig.~\ref{fig:contour_plots}, we plot the double differential cross section of the $\nu_\alpha A\to \nu_\alpha A \gamma$ process (for a $^{73}$Ge target) as a function of the outgoing photon energy $E_{\gamma}$ and angle $\theta_{\gamma}$ in the Dirac (left) and Majorana (right) cases, setting $\mathcal{F}(t_1) = 1$ for simplicity and choosing the benchmark values $E_\nu = 3$~MeV (approximately the peak of the reactor neutrino flux at Chooz) and $m_N = 1$~MeV (so that $N$ can be produced on-shell). In the plots, we choose the values $\mu^\alpha_{\nu N} = 3\times 10^{-8}$~$\mu_B$ and $\Gamma_N^{\text{inv}} \approx \Gamma_N = 10^{-11}$~MeV.
The difference between the Dirac and Majorana scenarios is striking; while both cases predict a considerable number of events at small photon energies, irrespective of the photon emission angle, there are more forward emissions of high energy photons in the Majorana case.

The single differential distributions in $E_\gamma$ and $\theta_\gamma$ are also different in the Dirac and Majorana scenarios. In the Dirac case the differential cross section in $E_\gamma$ decreases linearly from the minimum to maximum photon energies $E_\gamma^{-}$ and $E_\gamma^{+}$, respectively, while in the Majorana case the distribution is flat, i.e.
\begin{align}
	\label{eq:single_diff}
	\frac{d\sigma^{\text{D}}_{\nu_\alpha A\to XA\gamma}}{dE_\gamma} &= \frac{2\sigma^{\text{D}}_{\nu_\alpha A\to XA\gamma}(E_{\gamma}^{+}-E_{\gamma})}{(E_{\gamma}^{+}-E_{\gamma}^{-})^2}\,\Theta(E_\gamma-E_\gamma^{-})\Theta(E_\gamma^{+}-E_\gamma)\,,\\
	\frac{d\sigma^{\text{M}}_{\nu_\alpha A\to XA\gamma}}{dE_\gamma} &= \frac{\sigma^{\text{M}}_{\nu_\alpha A\to XA\gamma}}{E_{\gamma}^{+}-E_{\gamma}^{-}}\,\Theta(E_\gamma-E_\gamma^{-})\Theta(E_\gamma^{+}-E_\gamma)\,,
	\label{eq:single_diff2}
\end{align}
where $\sigma^{\text{M}}_{\nu_\alpha A\to XA\gamma} = 2\sigma^{\text{D}}_{\nu_\alpha A\to XA\gamma}$, $E_\gamma^{\pm} \approx \frac{E_\nu}{2}\big(1\pm\sqrt{1-\frac{m_N^2}{E_\nu^2}}\big)$, and $\Theta(x)$ is the Heaviside step function. Furthermore, the angular distribution in the lab frame peaks at slightly lower angles  in the Majorana case compared to the Dirac case. 

The lab-frame distributions are in exact agreement with the angular distributions in the rest frame of $N$, which can be derived purely from arguments of rotational and charge, parity and time reversal (CPT) invariance.  Due to the conservation of angular momentum, Dirac $N$ ($\bar{N}$) can only decay to left-polarised (right-polarised) photons $\gamma_{-}$ ($\gamma_+$) with an angular distribution in $\cos\vartheta_\gamma$ proportional to $(1+\cos\vartheta_\gamma)$ in the rest frame~\cite{Li:1981um, Shrock:1982sc, BahaBalantekin:2018ppj, Balantekin:2018ukw, Balaji:2019fxd, Balaji:2020oig, deGouvea:2021ual}. Majorana $N$ can decay equally to both left and right-polarised photons with angular distributions in $\cos\vartheta_\gamma$ proportional to $(1+\cos\vartheta_\gamma)$ and $(1-\cos\vartheta_\gamma)$, respectively; the total distribution is thus isotropic. The distinctive Dirac and Majorana energy distributions in Eqs.~\eqref{eq:single_diff} and \eqref{eq:single_diff2}, respectively, can be readily derived by boosting these rest-frame angular distributions to the lab frame. The photon circular polarisation provides an additional handle on the nature of the sterile neutrino and the CP properties of the transition dipole coupling (see, e.g., Ref.~\cite{Balaji:2019fxd}); a detailed analysis of the complementarity of a polarisation study is beyond the scope of this work and will be addressed elsewhere~\cite{Bolton:2021aaa}.

\section{Sensitivity to Active-Sterile Neutrino Magnetic Moments}
\label{sec:discussion}

In order to study the feasibility of our proposal, we will now examine the NUCLEUS experiment, which aims to detect CE$\nu$NS ($\bar{\nu}_e A\to \bar{\nu}_e A$) with nuclear recoil energies as low as $E_R\sim 10$~eV. Situated at the very-near-site (VNS) of the Chooz reactor site, the detector will receive an electron antineutrino flux of $\phi_{\bar{\nu}_e}\sim 10^{12}$~$\bar{\nu}_e$~cm$^{-2}$~s$^{-1}$. In the near future, phase~I of the experiment will use a 10~g $\mathrm{Al}_2 \mathrm{O}_3$/$\mathrm{CaWO}_4$ detector of size $L_{\text{det}}\sim 5$~cm, while in the far future, phase~II will upgrade to a 1~kg $^{73}$Ge detector of size $L_{\text{det}}\sim 25$~cm~\cite{NUCLEUS}. Interestingly, the detector will be sensitive to nuclear recoils and potentially to outgoing photons in the 1~keV to 10~MeV energy range. These can be detected at the cryogenic outer veto with an ionisation resolution of 50~--~100~keV for $\mathcal{O}$(MeV) photons~\cite{NUCLEUS}. A coincidence study between the nuclear recoil and an outgoing photon signal which can potentially lead to excellent background rejection. In this work, we therefore neglect any secondary backgrounds to the radiative mode as a first approximation.

For the Chooz reactor antineutrino flux, the maximum number of events are expected for sterile neutrino masses $m_N\sim 1\-- 5$~MeV. We find that the Dirac and Majorana cases show different differential rates in the photon energy, as expected from the differential cross sections shown in Fig.~\ref{fig:contour_plots}. The Majorana case presents a more symmetric distribution in the outgoing photon energy, while the Dirac case differential rate is shifted towards lower outgoing photon energies. For a specific sterile neutrino mass $m_N$, the cross section for the radiative upscattering process $\bar{\nu}_e A\to \bar{\nu}_e A\gamma$ grows with increasing incoming neutrino energy, with a resonant energy around $m_N$. Consequently, a different sterile neutrino mass can be probed for each neutrino energy in the Chooz spectrum. Average nuclear recoil and photon energies are increased for larger masses $m_N$. The total number of events falls off at low energies due to a smaller cross section, and at high energies due to the decreased flux. Since the number of events peaks at $E_\nu\sim m_N$, the decay width can be considered to be roughly constant. In this work, we only consider reactor antineutrinos as a proof-of-concept, though in principle a similar study could be performed for solar, atmospheric, or beam-dump neutrinos.

In order to identify the reach of the NUCLEUS experiment within the dipole portal parameter space, we consider both the Primakoff and radiative upscattering processes at the NUCLEUS detector. For the former process, the experiment will observe $N_{\text{obs}}$ nuclear recoils over its run time $T$. If the expected number of recoil events is given by $N_{\text{exp}} = N_{\text{bkg}} + N_{\nu A} + N_{N A}$, we can construct the chi-squared $\chi^2 = (N_\text{obs} - N_\text{exp})^2 / N_\text{exp}$. The nuclear recoil background has been estimated by the NUCLEUS collaboration to be $N_{\text{bkg}}=100~\text{keV}^{-1}~\text{kg}^{-1}~\text{day}^{-1}$~\cite{NUCLEUS:2019igx}. The number of recoil events induced by SM CE$\nu$NS, $N_{\nu A}$, and Primakoff upscattering, $N_{NA}$, are found by integrating the cross sections of these processes over the incoming energies of the Chooz $\bar{\nu}_e$ flux, from the minimum to maximum nuclear recoil energies, and multiplying by the detector mass and run time. We now assume that the experiment does not see an excess of recoil events over the SM CE$\nu$NS signal and background, i.e. $N_\text{obs} = N_{\text{bkg}} + N_{\nu A}$, giving $\chi^2 = N_{N A}^2/N_\text{exp}$. For simplicity, we do not include a nuisance parameter in $\chi^2$ to account for the presence of systematic errors. To set bounds at 90\% C.L., we find the allowed values of $\mu_{\nu N}^e$ satisfying $\chi^2 < 2.71$.

For the radiative upscattering process, the NUCLEUS experiment should in principle be able to detect $N^{\gamma}_{\text{obs}}$ coincidence events of a nuclear recoil accompanied by an outgoing photon. We assume a negligible SM background and therefore take the expected number of coincidence events, $N^\gamma_{\text{exp}} = N_{\nu A\gamma}$, to be Poisson distributed. Here, $N_{\nu A\gamma}$ is found by again integrating the radiative differential cross section in Eq.~\eqref{eq:cross_sec_NW} over the Chooz $\bar{\nu}_e$ flux, from the minimum to maximum nuclear recoil energies, and multiplying by the detector mass and operation time. To take into account the probability of the decay occurring within the detector, we must also multiply by the factor $P_{N}^{\text{det}}=1-\text{exp}(-\frac{L_{\text{det}} \Gamma_{N}}{\beta\gamma})$. Assuming that NUCLEUS does not observe any coincidence events, $N^{\gamma}_{\text{obs}}=0$, we set bounds at 90\% C.L. by finding the allowed values of $\mu_{\nu N}^e$ satisfying $N^\gamma_{\text{exp}} < 2.30$.

\begin{figure}[t!]
    \centering
    \includegraphics[width=0.9\textwidth]{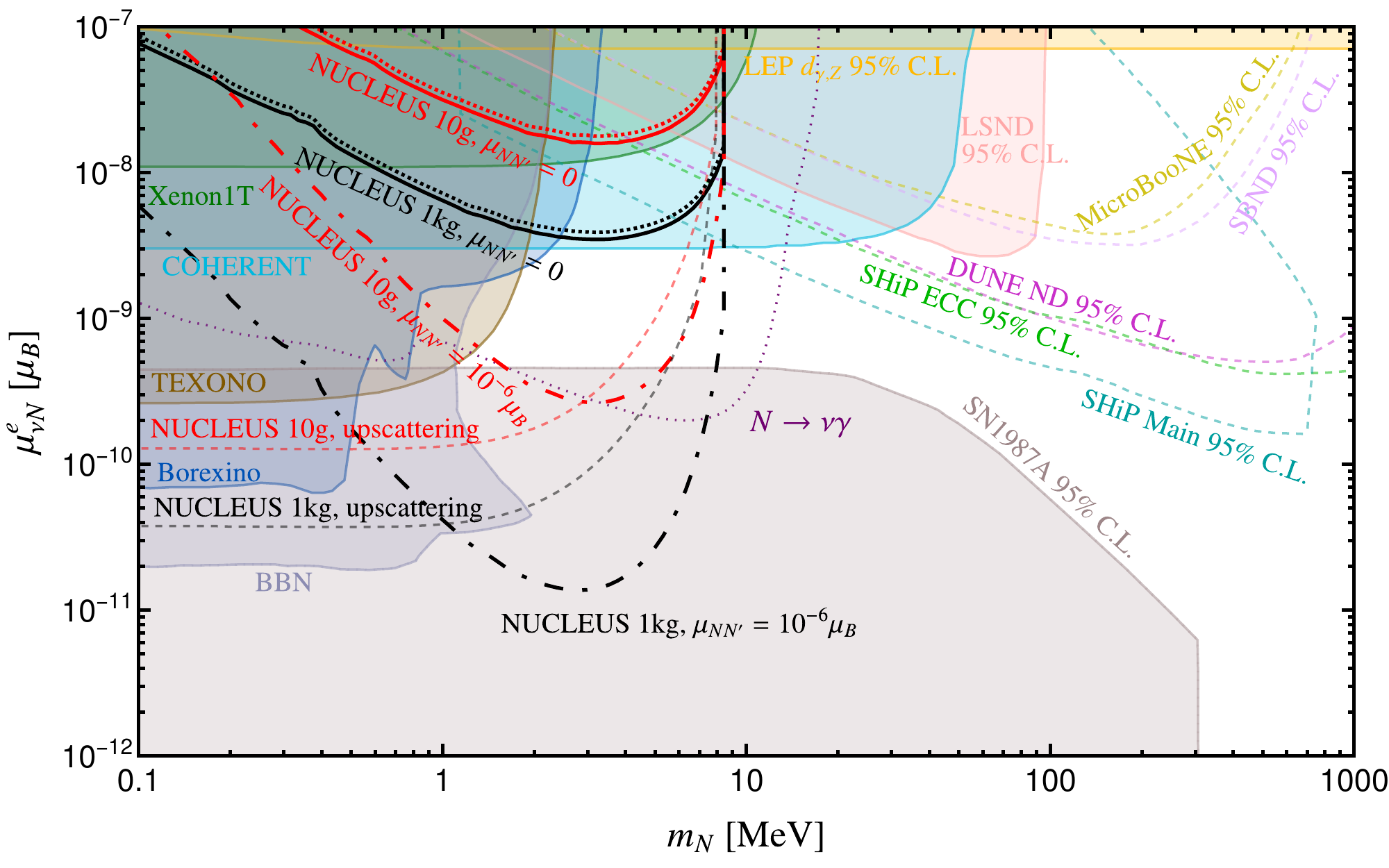}
    \caption{Constraints and sensitivities on the electron-flavour transition dipole coupling $\mu_{\nu N}^e$ as a function of the sterile neutrino mass $m_N$ from terrestrial experiments and astrophysical processes (solid lines, with excluded areas filled), as well as projected exclusion limits from future experiments (dashed lines). In black and red lines are the near-future (10~g $\mathrm{Al}_2 \mathrm{O}_3$/$\mathrm{CaWO}_4$, $L_{\text{det}}=5$~cm) and far-future (1~kg $^{73}$Ge, $L_{\text{det}}=25$~cm) projected sensitivities of the NUCLEUS experiment for Majorana $N$, respectively, using the coincidence of a nuclear recoil and an outgoing photon. The solid lines assume that $N$ can only decay via $N\to \nu_e\gamma$, with $\mu^e_{\nu N}\neq 0$. The dotted lines take $N$ to have additional invisible decay modes. The dot-dashed lines assume the presence of the additional decay channel $N\to N'\gamma$, with $m_{N'}\ll m_{N}$ and $\mu_{N'N}=10^{-6}~\mu_B$. The thin dashed lines also show the bounds the NUCLEUS experiment can make from the observation of just nuclear recoils. For the NUCLEUS limits we assume a run time of 2 years. Projections are at 90\% C.L. unless shown otherwise. The dotted purple line corresponding to $N\to\nu\gamma$ decay~\cite{Plestid:2020vqf} is only valid for the muon-flavour coupling $\mu_{\nu N}^\mu$ and is included for comparison.}
    \label{fig:nudp_constraints}
\end{figure}

In Fig.~\ref{fig:nudp_constraints}, we summarise the current and future constraints on the electron-flavour active-sterile dipole coupling $\mu^e_{\nu N}$ as a function of the sterile neutrino mass $m_N$. We show current bounds from terrestrial experiments~\cite{Magill:2018jla,Brdar:2020quo,Plestid:2020vqf,Schwetz:2020xra,Shoemaker:2018vii,Dasgupta:2021fpn,Coloma:2017ppo,Gninenko:1998nn,Atkinson:2021rnp} and astrophysical processes~\cite{Magill:2018jla,Brdar:2020quo,Coloma:2017ppo}  as solid lines (with excluded areas filled) and the expected sensitivities of future experiments as dashed lines. Using the chi-squared treatment above, we show as thin dashed red and black lines the near- and far-future sensitivities of the NUCLEUS experiment to the Primakoff upscattering. More precisely, these are the sensitivities of the current NUCLEUS detector (10~g of $\mathrm{Al}_2 \mathrm{O}_3$/$\mathrm{CaWO}_4$, $L_{\text{det}} = 5$~cm) and future upgrade (1~kg of $^{73}$Ge, $L_{\text{det}} = 25$~cm), assuming a run time of 2~years. We also show in cyan the current constraints derived in this work from Primakoff upscattering at the COHERENT experiment, noting a good agreement with Ref.~\cite{Miranda:2021kre}. 

The other red and black lines in Fig.~\ref{fig:nudp_constraints} are derived assuming the non-observation of a nuclear recoil and photon coincidence event in NUCLEUS. The solid lines depict the first benchmark scenario considered in Section~\ref{sec:calculation} in which $N$ can only decay via the active-sterile dipole coupling $\mu_{\nu N}^{e}$. While the radiative upscattering is suppressed by an additional factor of $|\mu_{\nu N}^{e}|^2$ with respect to the Primakoff upscattering, the negligible coincidence background results in similar sensitivities to the XENON1T and COHERENT experiments. The dotted red and blue lines instead depict the second benchmark scenario where $N$ has additional invisible decay modes with $\Gamma_{N}^{\text{inv}} = \beta\gamma/(5~\text{cm})\sim 10^{-11}$~MeV ($\Gamma_{N}^{\text{inv}} = \beta\gamma/(25~\text{cm})\sim 10^{-12}$~MeV) for the near-future (far-future) NUCLEUS phase. It can be seen that increasing $\Gamma_{N}^{\text{inv}}$ to this value does not appreciably impact the sensitivity; larger values of $\Gamma_{N}^{\text{inv}}$, however, result in weaker bounds. Finally, the dot-dashed lines show the reach of NUCLEUS in the third benchmark scenario in which $N$ decays predominantly to a lighter sterile state $N'$ via the sterile-sterile transition dipole coupling $\mu_{N'N} =  10^{-6}~\mu_B$. As discussed previously, the sterile-sterile dipole coupling $\mu_{N'N}$ is not subject to the same constraints as the active-sterile couplings $\mu^\alpha_{\nu N}$ (for example, those on $\mu^e_{\nu N}$ in Fig.~\ref{fig:nudp_constraints}). Consequently, the rate for the radiative upscattering process can be increased with respect to the Primakoff upscattering, leading to an improvement in sensitivity. For $m_N\gtrsim 1$~MeV, the near- and far-future sensitivities now constrain smaller values of $\mu_{\nu N}^e$ compared to the bounds from Primakoff upscattering.

\section{Conclusions}
\label{sec:conclusions}

For the radiative upscattering mode, a signal would consist of the coincidence of a nuclear recoil and an outgoing photon, separated by the decay length $\ell_N$ of the sterile state. In this work, we have proposed a novel approach in which the final state photons are searched for in a separate detector to the CE$\nu$NS target, and in particular we have studied the detection prospects of the reactor-based NUCLEUS experiment. As seen in Fig.~\ref{fig:nudp_constraints}, the current limits on the transition dipole coupling $\mu^e_{\nu N}$ derived from Primakoff upscattering at the COHERENT experiment are stringent, almost coinciding with sensitivity of the NUCLEUS experiment 1~kg upgrade in the radiative upscattering mode.

Using Primakoff upscattering, the 10~g NUCLEUS experiment will improve the sensitivity (red thin dashed), extending the limits to the region excluded by astrophysical observations for sterile neutrino masses $m_N\lesssim 10$~MeV. If the 10~g NUCLEUS experiment is indeed able to detect the Primakoff upscattering, it will provide an exciting motivation to search for the radiative upscattering mode in the 1~kg NUCLEUS upgrade. Even though the radiative mode is doubly suppressed by the dipole coupling as $|\mu^e_{\nu N}|^4$, its unique (potentially background-free) signature suggests that a future detection is not outside the realm of possibility. Such an observation would act as a smoking gun for our specific scenario and allow to differentiate it from other mechanisms.

The reach of the radiative upscattering mode can also be improved, as the intermediate sterile neutrino can decay via other transition magnetic moments, specifically with the $\nu_{\mu}$ and $\nu_{\tau}$ active neutrinos or other lighter sterile neutrinos $N'$. The radiative mode is thus generally proportional to $|\mu^e_{\nu N} \sum_X \mu_{X N}|^2$, with the sum over all lighter neutrinos in the final state. This scenario is indicated in Fig.~\ref{fig:nudp_constraints} by the dot-dashed curves for the 10~g (red) and 1~kg (black) NUCLEUS phases using a sterile-sterile transition dipole coupling $\mu_{N'N} = 10^{-6}~\mu_B$. For $m_N\gtrsim 1$~MeV, the limits are as stringent as those from Primakoff upscattering and remain competitive for $\mu_{N'N} \gtrsim 10^{-7}~\mu_B$.

Finally, an advantage of reactor-based CE$\nu$NS experiments is that they employ known fluxes of antineutrinos. If the radiative upscattering mode is observed, this opens up the possibility of discerning the Dirac or Majorana nature of the intermediate sterile neutrino solely with the energy and angular distribution of the photon, as demonstrated in Fig.~\ref{fig:contour_plots}.  This would have further implications on the neutrino mass generation mechanism, as well as theories of leptogenesis explaining the asymmetry between matter and antimatter in the universe. Specifically, if the sterile neutrino is found to be Majorana, the active neutrinos will also be of Majorana nature due to a mass term induced by the transition magnetic moment. We refer the interested reader to Appendix~\ref{sec:app_mass_bounds}, which provides a brief review of theoretical models in which transition magnetic moments are generated.


\begin{acknowledgments}
The authors are grateful to the NUCLEUS collaboration and in particular to B. Mauri, E. Mazzucato, C. Nones, J. Rothe, T. Lasserre,  F. Reindl, R. Strauss, M. Vignati for many useful discussions and communications regarding the experimental aspects of the NUCLEUS experiment. P.~D.~B. and F.~F.~D. acknowledge support from the UK Science and Technology Facilities Council (STFC) via the Consolidated Grants ST/P00072X/1 and ST/T000880/1. P.~D.~B. has received support from the European Union's Horizon 2020 research and innovation programme under the Marie Sk\l{}odowska-Curie grant agreement No 860881-HIDDeN. K.~F., J.~H. and C.~H. acknowledge support from the DFG Emmy Noether Grant No. HA 8555/1-1. K.~F. acknowledges partial support from the DFG Collaborative Research Centre “Neutrinos and Dark Matter in Astro- and Particle Physics” (SFB 1258). J.~H. acknowledges the fruitful discussions at the Aspen Center for Physics, which is supported by National Science Foundation grant PHY-1607611. S.~K. is supported by the Austrian Science Fund Elise-Richter grant project number V592-N27.
\end{acknowledgments}

\appendix

\section{Transition Dipole Moments and Implications for  Neutrino Masses}
\label{sec:app_mass_bounds}

\begin{figure}[t!]
	\centering
	\includegraphics[width=0.5\textwidth]{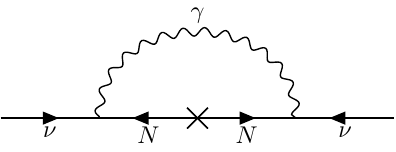}
	\caption{Loop contribution to a Majorana active neutrino mass from the transition magnetic dipole moment $\mu_{\nu N}$ and a Majorana sterile neutrino $N$.}
	\label{fig:maj_mass}
\end{figure}

In this appendix we comment on the possible implications of the discovery of an active to sterile transition dipole moment for active neutrino masses. If an experiment like NUCLEUS detects the differential distribution of the outgoing photon in the radiative upscattering process to be consistent with a heavy Majorana sterile state then that would necessarily imply that the light active neutrinos are Majorana. As can be easily noticed from Fig.~\ref{fig:maj_mass}, in the presence of an active to sterile transition dipole moment and a non-zero Majorana mass for the sterile state the active neutrinos will receive a Majorana mass contribution at the radiative level. While this radiative contribution may not necessarily account for the dominant contribution towards the active neutrino mass (making it dominantly Majorana), it necessarily implies a Majorana nature for the active neutrinos and hence violation of lepton number.  On the other hand, no conclusive remarks can be made about the nature of the active neutrino masses if the differential distribution of the radiative upscattering process turns out to be consistent with a Dirac sterile state, leaving both Dirac and Majorana possibilities open for the active neutrinos. Furthermore, a detection of radiative upscattering process mediated by a heavy Dirac or Majorana sterile state can also give interesting hints towards the mechanism of generating active neutrino mass. 

In many popular models of active neutrino mass generation the existence of an active to sterile transition dipole moment can be closely tied with a contribution to active neutrino masses. Therefore, the smallness of active neutrino masses can potentially disfavour many neutrino mass models  or render them to be \textit{unnatural} if the radiative upscattering process is observed. In fact, a large active to sterile transition dipole moment leading to an observable radiative upscattering rate together with smallness of active neutrino masses will hint towards a neutrino mass model with enhanced symmetry, e.g. a horizontal symmetry reinforcing the Voloshin mechanism~\cite{Voloshin:1987qy} or an inverse seesaw mechanism~\cite{Mohapatra:1986aw,Mohapatra:1986bd,Gonzalez-Garcia:1988okv}. Below we provide a brief overview of the relevant constraints and implications for Dirac and Majorana neutrino mass models. 

\textbf{Dirac neutrinos}: If the sterile state $N_R$ is a Weyl field with a coupling to the SM neutrino $\nu_L$ via the dipole interaction in Eq.~\eqref{eq:Ld}, then the active-sterile transition magnetic moments $\mu_{\nu N}$ can give rise to mass terms of the form $\mathcal{L} \supset m_{\nu N} \, \bar\nu_L N_R + \text{h.c.}$. This becomes more apparent when looked at from an effective field theory point of view as discussed in Ref.~\cite{Bell:2005kz}. An effective Lagrangian can be constructed as
\begin{align}
\label{eq:leff}
{\cal L}_{\rm eff} = \sum_{d,j} \frac{C^{(d)}_j(\mu)}{ \Lambda^{d-4}}\ {\cal O}^{(d)}_j(\mu) ~+~ {\rm h.c.}\,,
\end{align}
where the $d\geq 4$ denotes the operator dimension, $j$ runs over all independent operators of a given dimension, $\mu$ is the renormalisation scale, and $\Lambda$ corresponds to the new physics (NP) scale where new heavy degrees of freedom are integrated out. 

In a SM gauge group invariant theory an effective transition magnetic dipole moment can be generated by gauge-invariant, dimension-six ($d=6$) operators with couplings to the SU(2)$_L$ and U(1)$_Y$ gauge fields $W_\mu^a$ and $B_\mu$. Above the EW symmetry breaking scale, due to renormalisation group (RG) running these operators will mix with other $d=6$ operators that contain the SM lepton doublet $L=(\nu_L,e_L)^{T}$, $N_R$, and the SM Higgs field $H$. One such operator actually leads to generation of a Dirac neutrino mass term $m_{\nu N}$ after the EW symmetry breaking, as can be seen by considering the basis of independent operators at $d=6$ that are closed under renormalisation~\cite{Bell:2005kz}
\begin{eqnarray} \nonumber 
{\cal O}^{(6)}_1 & = & g_1{\bar L}{\tilde H}\sigma_{\mu\nu}N_R B^{\mu\nu} \,, \\
\label{eq:ops}
{\cal O}^{(6)}_2 & = & g_2 {\bar L}\tau^a {\tilde H} 
\sigma_{\mu\nu}N_R W^{a\mu\nu} \,, \\ \nonumber
{\cal O}^{(6)}_3 & = & {\bar L}{\tilde H}N_R \left(H^\dag H\right)\,, 
\end{eqnarray}
where $B_{\mu\nu}=\partial_\mu B_\nu-\partial_\nu B_\mu$ and
$W_{\mu\nu}^a=\partial_\mu W_\nu^a-\partial_\nu
W_\mu^a-g_2\epsilon_{abc}W_\mu^b W_\nu^c$ are the U(1)$_Y$ and
SU(2)$_L$ field strength tensors, respectively, $g_1$ and $g_2$
are the corresponding gauge couplings, and $\tilde{H}= i\sigma_2 H^*$. Starting from the Wilson coefficients $C^{(6)}_j(\mu=\Lambda)$ at the scale $\mu=\Lambda$, the RG running leads to mixings between ${\cal O}^{(6)}_{1,2}$ and ${\cal O}^{(6)}_3$ such that $C^{(6)}_3(\mu=v)$ receives a contribution from $C^{(6)}_{1,2}(\mu=\Lambda)$~\cite{Bell:2005kz}. After the EW symmetry breaking the combination $C^{(6)}_1{\cal O}^{(6)}_1+C^{(6)}_2 {\cal O}^{(6)}_2$ leads to the magnetic moment
\begin{align}
\label{eq:Dmunu1}
\frac{\mu_{\nu N}}{\mu_B}=
-16\sqrt{2}\left( \frac{m_e v}{\Lambda^2}\right) \left[C^{(6)}_1(v)+C^{(6)}_2(v)\right]\,,
\end{align}
where $\mu=v$ corresponds to the EW symmetry breaking scale. On the other hand ${\cal O}^{(6)}_3$ leads to the Dirac mass term
\begin{align}
\label{eq:Ddeltamnu1}
\delta m_{\nu N} = -C^{(6)}_3(v) \frac{v^3}{2\sqrt{2}\Lambda^2}\,,
\end{align}
yielding a relation between $\delta m_{\nu N}$ and $\mu_{\nu N}$ given by
\begin{equation}
\label{eq:munurel}
\delta m_{\nu N}=\frac{v^2}{16 m_e}\ \frac{C^{(6)}_3(v)}{C^{(6)}_1(v)+C^{(6)}_2(v)}\ \frac{\mu_{\nu N}}{\mu_B}\,,
\end{equation}
which leads to the constraint
\begin{equation}
\label{eq:massbound}
\frac{|\mu_{\nu N}|}{\mu_B} \sim 10^{-15}\left(\frac{\delta m_{\nu N}}{1\ {\rm eV}}\right)\,,
\end{equation}
for $\Lambda=1$ TeV. This implies that for a realistic active neutrino Dirac mass $m_\nu \lesssim 1$ eV, $|\mu_{\nu N}|\lesssim 10^{-15}\mu_B$. 

However, we note that this constraint is strictly applicable when $N_R$ is a Weyl field forming a Dirac pair with $\nu_L$. The above constraint does not hold true in a number of general circumstances. One example is the scenario in which $N$ is a Dirac fermion containing two Weyl fields (i.e. $N = N_R + N_L$) and has a Dirac mass $m_N \gg m_\nu^{\text{obs}}$, which can \textit{a priori} be completely decoupled from the generation of the active neutrino masses. This is the Dirac scenario we consider in this work, and therefore the constraint in Eq.~\eqref{eq:massbound} can be safely neglected. Another example is the scenario in which the tree-level Dirac mass term $m_{\nu N}^{\text{SM}}$ is of the same order but of opposite sign compared to the $\delta m_{\nu N}$ generated by $\mu_{\nu N}$. While some might consider this cancellation to be unnatural, it is by no means improbable. A third example is a Dirac seesaw scenario where $N$ is a Dirac fermion (possibly vector-like) and with a Dirac mass $m_N \gg m_\nu^{\text{obs}}$ and the active neutrino $\nu_L$ forms a Dirac pair with a new SM gauge-singlet Weyl field $\nu_R$, which can have a Dirac mass term of the form $m_{\nu_R N} \, \bar N_L \nu_R$ (which can arise naturally from the vacuum expectation value of a SM gauge-singlet scalar state). In such a situation the active neutrino can receive a \textit{purely Dirac} mass contribution of the form $m_\nu \sim m_{\nu N} m_{\nu_R N}/m_N$~\cite{Cepedello:2018zvr,Bolton:2019bou}.

\textbf{Majorana neutrinos}: In the case of the type~I seesaw scenario, the presence of a transition magnetic dipole moment via a loop diagram with heavy NP in the loop also leads to a contribution to Dirac mass term of the from $\mathcal{L} \supset m_{\nu N} \, \bar\nu_L N_R$, through the same diagram but with the external photon line removed. This leads to a naive relation between $\mu_{\nu N}$ and the NP loop-induced $\delta m_{\nu N}$,
\begin{align} \label{eq:maj_con_1}
\frac{\mu_{\nu N}}{\mu_B} \approx \frac{m_e \delta m_{\nu N}}{\Lambda^2}\,.
\end{align}
However, one can always tune the tree-level Yukawa coupling $\mathcal{L}_Y \supset y_\nu \, \bar L_L \tilde{H} N_R+\text{h.c.}$ such that the resulting tree-level contribution to the Dirac mass, $y_\nu v/\sqrt{2}$, nearly cancels the loop-induced contribution. In this way, one can lower the right-handed neutrino masses $m_N$ to as low as MeV scales while keeping the active neutrino masses at $m_\nu \lesssim 1$~eV. 

Exceptions to the relation in Eq.~\eqref{eq:maj_con_1} can naturally occur when additional symmetries are present. An example is the so-called Voloshin mechanism, where an approximate global $SU(2)_H$ symmetry is introduced such that $(\nu_L^{c}, N_R)$ transforms as a doublet under the $SU(2)_H$~\cite{Voloshin:1987qy}. While this symmetry allows for a $SU(2)_H$
singlet transition magnetic moment term of the form $\bar{N}_R \sigma^{\mu \nu} \nu_L -
\bar{\nu}_L^{c} \sigma^{\mu \nu} N_R^{c}$, it naturally forbids the $SU(2)_H$ triplet contribution to the neutrino mass term $\bar{N}_R \nu_L + \bar{\nu}_L^{c} N_R^{c}$. Instances of recent models employing the Voloshin mechanism in the context of neutrino magnetic dipole moments can be found in Refs.~\cite{Brdar:2020quo,Babu:2020ivd}.

In the presence of a sizeable mass mixing between active and sterile states (as can be the case in a typical type~I seesaw) an active-sterile transition magnetic dipole moment can also be induced through loop diagrams involving charged leptons~\cite{Pal:1981rm,Shrock:1982sc}. Such an active-sterile mass mixing induced contribution to the transition magnetic dipole moment is given by~\cite{Magill:2018jla}
\begin{equation}\label{eq:maj_con_2}
\begin{split}
\frac{|\mu_{\nu N}|}{\mu_B}=\frac{3m_{\nu N} m_e}{16\pi^2}\frac{G_{F}}{\sqrt{2}}
\sim  10^{-13}\left(\frac{m_{\nu N}}{1~\text{MeV}}\right)\,.
\end{split}
\end{equation}

On the other hand, in the presence of a transition magnetic moment between $\nu_L$ and $N$, a loop contribution to the light active neutrino masses is induced through Fig.~\ref{fig:maj_mass}, which is directly proportional to the Majorana mass of $N$,
\begin{align} \label{eq:maj_con_3}
m_{\nu} \sim \left(\frac{\mu_{\nu N}}{\mu_B}\right)^2 \frac{\alpha}{16\pi} \frac{m_N \Lambda^2}{m_e^2}\,,
\end{align}
where $\Lambda$ is the cutoff scale for the UV completion of the model. For $m_N \sim 1$ MeV, $\Lambda\sim 1$ TeV and $m_{\nu}\lesssim 1$ eV, Eq.~\eqref{eq:maj_con_3} leads to $\frac{|\mu_{\nu N}|}{\mu_B}< 10^{-8}$. While a larger Majorana mass of $N$ would lead to relaxation of the tight constraint from Eq.~\eqref{eq:maj_con_2}, it would make the constraint from Eq.~\eqref{eq:maj_con_3} more stringent. However, one can easily circumvent these constraints by considering a scenario with a quasi-Dirac $N$, such as in the case of the inverse seesaw mechanism~\cite{Mohapatra:1986aw,Mohapatra:1986bd,Gonzalez-Garcia:1988okv}, where an approximate lepton number conservation (due to a very small Majorana mass splitting of a predominantly Dirac $N$ pair) makes the active neutrinos very light, while the $N$ being pseudo-Dirac can be significantly heavier.

\section{Calculation of the Radiative Upscattering Process}
\label{sec:app_calc}

In this appendix we derive the differential cross section for the radiative upscattering process $\nu_\alpha A \to X A \gamma$, $X=\{\nu_\beta, N',\ldots\}$. For completeness, we also derive the differential cross section for the Primakoff upscattering process $\nu_\alpha A \to N A$.

Firstly, we note that the active and sterile neutrinos $\nu_\alpha$ and $N$ ($N'$) may either be Dirac or Majorana fermions. Rates for the Primakoff and radiative upscattering processes can be therefore computed for the four possible combinations of Dirac or Majorana $\nu_\alpha$ and $N$ ($N'$). We will see that in the limit of massless neutrinos in the initial and final states, i.e. $m_\nu \to 0$ and $m_{N'} \to 0$, the rates for processes with Dirac or Majorana $\nu_\alpha$ (and $N'$) are identical, in accordance with the \textit{practical Dirac-Majorana confusion theorem}~\cite{Kayser:1981nw,Kayser:1982br}. However, for massive $m_N$, there is indeed a distinction between the rates for the $\nu_\alpha A \to X A \gamma$ process when $N$ is a Dirac or Majorana fermion.

\begin{figure}
	\centering
	\includegraphics[width=6cm]{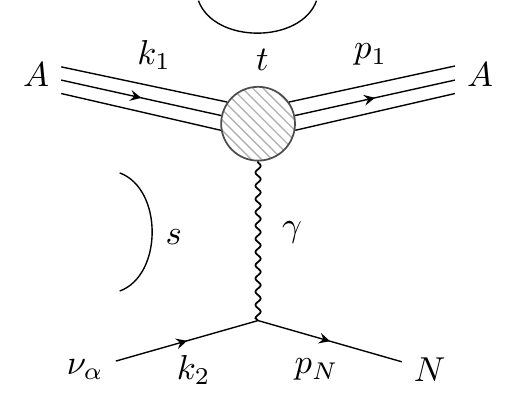}
	\includegraphics[width=6.8cm]{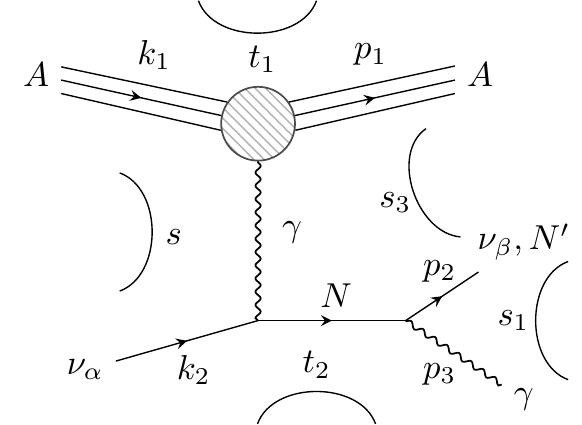}
	\caption{Diagrams for the Primakoff upscattering $\nu_\alpha A \to N A$ (left) and radiative scattering $\nu_\alpha A \to X A \gamma$ (right), $X=\{\nu_\beta,N',\ldots\}$, with momentum and Mandelstam variable assignments defined in the main text.}
	\label{fig:feyndiag2}
\end{figure}

In both the Dirac and Majorana cases, the transition magnetic moment between the light active neutrinos $\nu_\alpha$ (or a sterile neutrino $N'$) and the sterile neutrino $N$ is described by the effective Lagrangian
\begin{align}
\label{eq:LdSM}
\mathcal{L} \supset \frac{\mu_{X N}}{2}\bar{X} \sigma_{\mu\nu} P_R N F^{\mu\nu} + \frac{(\mu_{X N})^*}{2}\bar{N}\sigma_{\mu\nu}P_L X F^{\mu\nu}\,,
\end{align}
where $X=\{\nu_{\alpha L}, N',\ldots\}$ and $\mu_{X N}=\{\mu^\alpha_{\nu N}, \mu_{N' N},\ldots\}$. In the case where both $N$ and $X$ are Majorana, the second term can be rewritten to give
\begin{align}
\label{eq:MajLagrangian}
\mathcal{L} &\supset \frac{\mu_{X N}}{2}\bar{X} \Big[\sigma_{\mu\nu} P_R - \mathcal{C}(\sigma_{\mu\nu} P_L)^{T}\mathcal{C}^{-1} \Big] N F^{\mu\nu} = \frac{\mu_{X N}}{2} \bar{X}\sigma_{\mu\nu} N  F^{\mu\nu}
\end{align}
where we have defined $(\mu_{X N})^* = -\mu_{X N}$ and used the charge-conjugation properties $\mathcal{C}P^{T}_L\mathcal{C}^{-1} = P_L$ and $\mathcal{C}\sigma_{\mu\nu}^{T}\mathcal{C}^{-1} = -\sigma_{\mu\nu}$. If $X$ is Majorana and $N$ is Dirac (or vice versa) it is not possible to make such a simplification; it is nonetheless convenient to write in the first scenario (Majorana $X$, Dirac $N$),
\begin{align}
\label{eq:MajDirac1}
\mathcal{L} &\supset \frac{\mu_{X N}}{2} \bar{X}\Big[\sigma_{\mu\nu} P_R N -\mathcal{C}(\sigma_{\mu\nu} P_L)^{T}\mathcal{C}^{-1}N^c\Big]  F^{\mu\nu},
\end{align}
while in the latter case (Dirac $X$, Majorana $N$),
\begin{align}
\label{eq:MajDirac2}
\mathcal{L} &\supset \frac{\mu_{X N}}{2}\Big[\bar{X}\sigma_{\mu\nu} P_R - \bar{X}^c\mathcal{C}(\sigma_{\mu\nu} P_L)^{T}\mathcal{C}^{-1} \Big] N F^{\mu\nu}.
\end{align}
In the following we will use these interaction terms to investigate the Primakoff and radiative upscattering processes for Dirac and Majorana $\nu_\alpha$, $N$ and $N'$.

We finally note that if $X$ and $N$ are both Dirac, the following effective Lagrangian terms can also be written
\begin{align}
\label{eq:LdSM2}
\mathcal{L} \supset \frac{\zeta_{X N}}{2}\bar{X} \sigma_{\mu\nu} P_L N F^{\mu\nu} + \frac{(\zeta_{X N})^*}{2}\bar{N}\sigma_{\mu\nu}P_R X F^{\mu\nu}\,.
\end{align}
However, as we are considering purely left-handed (right-handed) incoming neutrinos (antineutrinos), these terms do not contribute to the Primakoff or radiative upscattering processes.

\textbf{Primakoff upscattering}: The active-sterile transition magnetic moment $\mu^\alpha_{\nu N}$ induces the Primakoff upscattering process shown to the left in Fig.~\ref{fig:feyndiag2}. An incoming Dirac or Majorana active neutrino $\nu_\alpha$ exchanges a photon with a target nucleus $A$ and scatters to an outgoing Dirac or Majorana sterile neutrino~$N$. As shown in the diagram, the ingoing nucleus and neutrino and outgoing nucleus and sterile neutrino have four-momenta $k_1$, $k_2$, $p_1$ and $p_N$, respectively. The four-momentum exchanged by the photon is thus $q = k_1 - p_1 = p_N - k_2$.

We first consider the scenario where $N$ is a Dirac fermion, i.e. $N = N_L + N_R$. An incoming neutrino $\nu_\alpha$, created by the SM charged current $j_{W}^\mu = \bar{\nu}_{\alpha L} \gamma^\mu \ell_{\alpha L}$, may be a Dirac or Majorana fermion. If $\nu_\alpha$ is Dirac, the neutrino is annihilated by the second term in Eq.~\eqref{eq:LdSM}. This results in the following matrix element for $\nu_\alpha A\to N A$,
\begin{align}
\label{eq:upscatteringmatrix}
i\mathcal{M}^{\text{D}}_{\nu_\alpha A\to N A} &= (\mu^\alpha_{\nu N})^*\left[\bar{u}_N\sigma_{\rho\sigma}P_{L}q^{\sigma}u_{\nu_\alpha}\right]\frac{(-ig^{\rho\lambda})}{q^2}\mathcal{J}^{A}_{\lambda}\,,
\end{align}
where $\mathcal{J}^{A}_{\lambda}$ is the hadronic current of the nucleus. In the following we use the hadronic current $\mathcal{J}^{A}_{\lambda} = -ieZ(\bar{u}_A\gamma_{\lambda} u_A)\mathcal{F}(q^2)$, where $\mathcal{F}(q^2)$ is a nuclear form factor, which describes the electromagnetic interaction of a spin-$\frac{1}{2}$ nucleus. If $\nu_\alpha$ is Majorana, the neutrino created by the SM charged current $j_{W}^\mu = \bar{\nu}_{\alpha L} \gamma^\mu \ell_{\alpha L}$ can instead be annihilated by the first or second term in Eq.~\eqref{eq:MajDirac1}. The second term corresponds to the propagation of a negative helicity neutrino and is again described by the matrix element in Eq.~\eqref{eq:upscatteringmatrix}. The first term on the other hand corresponds to the propagation of a positive helicity neutrino; because the neutrino is ultrarelativistic, this process is helicity suppressed by the small ratio $m_\nu/E_\nu$. If $\nu_\alpha$ is Dirac, the SM charged current $j_{W}^{\mu\dagger} = \bar{\ell}_{\alpha L} \gamma^\mu \nu_{\alpha L}$ instead creates an antineutrino $\bar{\nu}_\alpha$, which is annihilated by the first term in Eq.~\eqref{eq:LdSM}. The matrix element is
\begin{align}
\label{eq:upscatteringmatrix2}
i\mathcal{M}^{\text{D}}_{\bar{\nu}_\alpha A\to \bar{N} A} &= \mu^\alpha_{\nu N}\left[\bar{v}_{\nu_\alpha}\sigma_{\rho\sigma}P_{R}q^{\sigma}v_{N}\right]\frac{(-ig^{\rho\lambda})}{q^2}\mathcal{J}^{A}_{\lambda}\,.
\end{align}
If $\nu_\alpha$ is Majorana, the neutrino created by the charged current $j_{W}^{\mu\dagger} = \bar{\ell}_{\alpha L} \gamma^\mu \nu_{\alpha L}$ can again be annihilated by the first or second term in Eq.~\eqref{eq:MajDirac1}. The first term induces the process $\nu_\alpha A\to \bar{N} A$ with the matrix element in Eq.~\eqref{eq:upscatteringmatrix2}. The second term induces the process $\nu_\alpha A\to N A$ but is helicity suppressed by $m_{\nu}/E_\nu$.

We next consider the scenario where $N$ is a Majorana fermion, i.e. $N = N^c_R + N_R$. If the incoming neutrino $\nu_\alpha$ produced by the charged current $j_{W}^\mu = \bar{\nu}_{\alpha L} \gamma^\mu \ell_{\alpha L}$ is Dirac, it can only be annihilated by the second term in Eq.~\eqref{eq:MajDirac2}. This induces the process $\nu_\alpha A\to N A$ with the matrix element identical to Eq.~\eqref{eq:upscatteringmatrix}. Similarly, an antineutrino created by the charged current $j_{W}^{\mu\dagger} = \bar{\ell}_{\alpha L} \gamma^\mu \nu_{\alpha L}$ can only be annihilated by the first term in Eq.~\eqref{eq:MajDirac2}. This induces the process $\bar{\nu}_\alpha A\to N A$ with a matrix element identical to Eq.~\eqref{eq:upscatteringmatrix2}. An incoming Majorana neutrino $\nu_\alpha$ created by $j_{W}^\mu = \bar{\nu}_{\alpha L} \gamma^\mu \ell_{\alpha L}$ can be annihilated by both terms in ~\eqref{eq:MajLagrangian}. However, the contribution to the process $\nu_\alpha A\to N A$ from the first term is helicity suppressed. Likewise, an incoming Majorana neutrino $\nu_\alpha$ created by $j_{W}^{\mu\dagger} = \bar{\ell}_{\alpha L} \gamma^\mu \nu_{\alpha L}$ can also be annihilated by both terms in Eq.~\eqref{eq:MajLagrangian}, but the contribution from the second term is suppressed. Consequently, the matrix elements for these processes are also given by Eqs.~\eqref{eq:upscatteringmatrix} and \eqref{eq:upscatteringmatrix2}, respectively.

The differential cross section for the process can now be found by taking the absolute square of the scattering amplitude, averaging over the spin of the incoming nucleus and summing over the spins of the outgoing nucleus and sterile neutrino. Neglecting the mass of the incoming neutrino, the differential cross section is calculated as
\begin{align}
\label{eq:sigmaup}
d^2\sigma_{\nu_\alpha A\to N A} &= \frac{1}{2(s-m_A^2)}\frac{1}{2}\sum_{\mathrm{spins}}|\mathcal{M}^{\text{D}(\text{M})}_{\nu_\alpha A\to N A}|^2 d\Phi_2 \,,
\end{align}
where $s = (k_1 + k_2)^2$ is a Mandelstam variable, $m_A$ is the mass of the nucleus, and $d\Phi_2$ is the two-body phase space of the outgoing nucleus and sterile neutrino,
\begin{align}
\label{eq:2bodyphasespace}
d\Phi_2 = (2\pi)^4 \frac{d^3 \mathbf{p}_1}{2(2\pi)^3E_{\mathbf{p}_1}}\frac{d^3 \mathbf{p}_N}{2(2\pi)^3E_{\mathbf{p}_N}}\delta^{4}(k_1+k_2-p_1-p_N) =\frac{d\phi}{2\pi}\frac{dt}{8\pi(s-m_A^2)}\,.
\end{align}
In the second equality we have used the Dirac delta function (enforcing the conservation of four-momentum) to integrate over four of the three-momentum components. We have also re-expressed the integral in terms of $\phi$ (the azimuthal angle defining rotations around the incoming neutrino direction) and the Mandelstam variable $t = q^2 =(k_1 - p_1)^2$. The physical region of the phase space is defined by the inequality $\Delta_3 < 0$, where
\begin{align}
\Delta_3 = -\frac{1}{2}\large\left|\begin{smallmatrix}
0 & 0 & s & m_N^2 & 1 \\
0 & 0 & m_A^2 & t & 1 \\
s & m_A^2 & 0 & m_{A}^2 & 1 \\
m_N^2 & t & m_A^2 & 0 & 1 \\
1 & 1 & 1 & 1 & 0 \\
\end{smallmatrix}\right|\,
\end{align}
is the $3\times 3$ symmetric Gram determinant constructed from any three of the four incoming or outgoing four-momenta~\cite{Byckling:1971vca}.
The spin-averaged and summed squared matrix elements for the process $\nu_\alpha A\to N A$ can be written, for both Dirac and Majorana $N$, as
\begin{align}
\label{eq:upscatteringMsq}
\frac{1}{2}\sum_{\mathrm{spins}}|\mathcal{M}_{\nu_\alpha A\to N A}|^2 &= |\mu^\alpha_{\nu N}|^2 \,\frac{e^2 Z^2\mathcal{F}^2(q^2)}{q^4}L_{\mu\nu}H^{\mu\nu}\,,
\end{align}
where the leptonic and hadronic components are given by
\begin{align}
\label{eq:Lmunu}
L_{\mu\nu} &= \mathrm{Tr}[(\slashed{p}\hspace{-0.5em}\phantom{p}_N+m_N)\sigma_{\mu\rho}P_{L}\slashed{k}_2\sigma_{\nu\lambda}]\,q^{\rho}q^{\lambda}\,, \\
H^{\mu\nu} &= \frac{1}{2}\,\mathrm{Tr}[(\slashed{p}\hspace{-0.5em}\phantom{p}_1+m_A)\gamma^{\mu}(\slashed{k}\hspace{-0.5em}\phantom{k}_1+m_A)\gamma^{\nu}]\,.
\label{eq:Wmunu}
\end{align}
As mentioned previously, the form of the hadronic current $H^{\mu\nu} $ in Eq.~\eqref{eq:Wmunu} is technically only valid for spin-$\frac{1}{2}$ nuclei, while most targets considered for CE$\nu$NS are spin-0. However, for small recoil energies, which is the regime of interest for most CE$\nu$NS experiments, the spin of the nucleus has a negligible impact on the cross section. We can now insert Eqs.~\eqref{eq:2bodyphasespace} and \eqref{eq:upscatteringMsq} into Eq.~\eqref{eq:sigmaup} and express the matrix element squared in terms of the Mandelstam variables $s$ and $t$. The matrix element squared does not depend on the azimuthal angle $\phi$, which can therefore be integrated over.  This gives the following differential cross section in the Lorentz-invariant variable $t$,
\begin{align}
\label{eq:sigmaupt}
\frac{d\sigma_{\nu_\alpha A\to N A}}{dt} &=  |\mu^\alpha_{\nu N}|^2\alpha Z^2\mathcal{F}^2(t)\frac{F(t,s,m_A^2,m_N^2)}{2t^2(s-m_A^2)^2}\,,
\end{align}
where the function $F(a,b,c,d) = -2 a(b - c)^2 -  2a^2(b - c) + a d(a + 2 b) - d^2(a + 2c)$ has been introduced for convenience.

We would now like to determine the cross section as a function of lab frame variables. In the lab frame, $k_1 = (m_A,\mathbf{0})$, $k_2 = (E_\nu,\mathbf{k}_2)$, $p_1 = (m_A+E_R,\mathbf{p}_1)$ and $p_N = (E_\nu-E_R,\mathbf{p}_N)$, where $E_\nu$ is the incoming neutrino energy and $E_R$ is the nuclear recoil energy. In the lab frame the Mandelstam variables are $s = m_A(m_A+2E_\nu)$ and $t = -2 m_A E_R$. We transform the variable from $t$ to $E_R$ by multiplying by the Jacobian factor $\partial t/\partial E_R = -2 m_A$, which gives the standard result
\begin{align}
\label{eq:sigmalabframe}
\frac{d\sigma_{\nu_\alpha A\to N A}}{dE_R} = |\mu^\alpha_{\nu N}|^2\alpha Z^2\mathcal{F}^2(E_R)\bigg[\frac{1}{E_{R}}-\frac{1}{E_{\nu}}&-\frac{m_{N}^2}{4E_{R}E_{\nu}^2}\left(1+\frac{2E_{\nu}-E_R}{m_A}\right) \nonumber\\
&-\frac{m_{N}^4}{8m_{A}E_{R}^2E_{\nu}^2}\left(1-\frac{E_R}{m_A}\right)\bigg]\,.
\end{align}
In this work, we are interested in the limit in which the momentum exchanged by the photon (and therefore the nuclear recoil $E_R$) is much smaller than the mass of the nucleus $m_A$. We also assume that the sterile neutrino mass $m_N$ is comparible to the incoming neutrino energy $E_\nu$, but much smaller than $m_A$. In terms of the Mandelstam variables this corresponds to $s\approx m_A^2\gg t, m_{N}^2$, and the function in Eq.~\eqref{eq:sigmaupt} simplifies to
$F(s,t,m_A^2,m_N^2) \approx -2t(s-m_A^2)^2 + 2 m_A^2 m_N^2 (t-m_N^2)$. This is equivalent to dropping the terms proportional to $1/E_{\nu}$ and $1/m_A$ in Eq.~\eqref{eq:sigmalabframe}.

\textbf{Radiative upscattering}:
We now outline how to compute the matrix element squared and differential cross section of the radiative upscattering process $\nu_\alpha A\to X A \gamma$, i.e. the Primakoff upscattering $\nu_\alpha A\to N A$ followed by the radiative decay $N \to X \gamma$, where $X$ can either be an active neutrino $\nu_\beta$ or another sterile state $N'$.

Before doing so, it is useful to derive the decay rate for the process $N\to X \gamma$ induced by the transition magnetic dipole moment $\mu_{X N}$. If $N$ and $X$ are Dirac fermions, sterile neutrinos $N$ and antineutrinos $\bar{N}$ decay to $X$ and $\bar{X}$, respectively. The matrix element for $N\to X \gamma$ is
\begin{align}
\label{eq:gammadecaymatrix}
i\mathcal{M}^{\text{D}}_{N\to X \gamma} &= \mu_{X N} [\bar{u}_{X} \sigma_{\mu\nu} P_R u_N] \epsilon^{\mu*} p^\nu_3\,,
\end{align}
where $p_3$ and $\epsilon$ are the four-momentum and polarisation of the outgoing photon, respectively. The matrix element for $\bar{N}\to \bar{X}\gamma$ is determined from the second term in Eq.~\eqref{eq:LdSM} and is given by Eq.~\eqref{eq:gammadecaymatrix} with $\mu_{XN}\to(\mu_{XN})^*$ and $[\bar{u}_{X} \sigma_{\mu\nu} P_R u_N] \to [\bar{v}_{N} \sigma_{\mu\nu} P_L v_{X}]$. The same expressions are valid for Dirac sterile neutrinos $N$ and antineutrinos $\bar{N}$ decaying to Majorana $X$ as they can only be annihilated by the first and second terms in Eq.~\eqref{eq:MajDirac1}, respectively.

We now examine the case if $N$ is instead a Majorana fermion. If $X$ is Dirac, $N$ can decay to $X$ and $\bar{X}$ via the first and second terms of Eq.~\eqref{eq:MajDirac2}, respectively. If $X$ is Majorana, $N$ can decay via both of the terms in Eq.~\eqref{eq:MajLagrangian}. The matrix element for the sum of processes $N\to X\gamma$ and $N\to \bar{X}\gamma$ for Dirac $X$ (and $N\to X\gamma$ for Majorana $X$) is thus given by
\begin{align}
\label{eq:gammadecaymatrixmaj}
i\mathcal{M}^{\text{M}}_{N\to X \gamma} &= \mu_{X N} \big[\bar{u}_{X} \big(\sigma_{\mu\nu} P_R -  \mathcal{C}(\sigma_{\mu\nu} P_L)^{T}\mathcal{C}^{-1}\big) u_N \big] \epsilon^{\mu*} p^\nu_3  \nonumber\\
&=\mu_{X N} [\bar{u}_{X} \sigma_{\mu\nu}  u_N] \epsilon^{\mu*} p^\nu_3\,.
\end{align}
To compute the decay rate, we multiply the spin- and polarisation-summed squared matrix element by the two-body phase space of the outgoing neutrino and photon,
\begin{align}
\label{eq:gammadecay}
d^2\Gamma^{\text{D}(\text{M})}_{N\to X \gamma} &= \frac{1}{2m_N}\sum_{\mathrm{spins},\,\mathrm{pols}}|\mathcal{M}^{\text{D}(\text{M})}_{N\to X \gamma}|^2 d\Phi_2 \,.
\end{align}
As we will later be considering the sterile neutrino $N$ as an intermediate particle in the scattering process $\nu_\alpha A\to X A \gamma$, we do not average over its spin. It is straightforward to find (neglecting the mass of $X$)
\begin{align}
\sum\limits_{\text{spins},\,\text{pols}} |\mathcal{M}^{\text{M}}_{N\to X \gamma}|^2 = 2\sum\limits_{\text{spins},\,\text{pols}}|\mathcal{M}^{\text{D}}_{N\to X \gamma}|^2 = 4  m_N^4 |\mu_{X N}|^2\,,
\end{align}
where we note the factor of two difference between the Dirac and Majorana case. Writing $d\Phi_2$ in terms of the angles of the outgoing neutrino and photon, i.e. $d\Phi_2 = \frac{1}{8\pi}\frac{d\phi}{2\pi}\frac{d\cos\theta}{2}$, we see that the form of the squared matrix elements allows to integrate over $\phi$ and $\cos\theta$, giving
\begin{align}
\label{eq:decay_rate}
\Gamma^{\text{M}}_{N\to X\gamma} = 2\Gamma^{\text{D}}_{N\to X\gamma} = \frac{m_N^3|\mu_{X N}|^2}{4\pi}\,.
\end{align}
The Majorana decay rate via the transition magnetic moment $\mu_{X N}$ is a \textit{factor of two} larger than the corresponding Dirac decay rate. 


We now return to the radiative upscattering process $\nu_\alpha A\to X A \gamma$. The Feynman diagram for this process is shown to the right of Fig.~\ref{fig:feyndiag2}, where we define the momenta of the incoming and outgoing particles and the Mandelstam variables $s$, $s_1$, $t_1$, $s_3$ and $t_2$. The variables $s$ and $t_1 = q^2$ are equivalent to $s$ and $t$ for the $\nu_\alpha A\to N A$ process. In the lab frame, these are
\begin{align}
\label{eq:sMandelstam}
s & = (k_1 + k_2)^2 = m_A(m_A+2E_\nu)\,, \\
t_1 &= (k_1 - p_1)^2 = -2 m_A E_R\,, \\
s_1 & =(p_2+p_3)^2 = -2 m_A E_R - 2 E_\nu (E_R - \sqrt{E_R (2 m_A + E_R)} \cos\theta_R)\,, \label{eq:s1Mandelstam}\\
s_3 & = (p_1+p_2)^2=  m_A (m_A + 2E_\nu - 2E_\gamma)
-2 E_\nu E_\gamma (1 - \cos\theta_\gamma)\,,\\
t_2 &= (k_2 - p_3)^2 = - 2 E_\nu E_\gamma  (1-\cos\theta_\gamma)\,,
\label{eq:t2Mandelstam}
\end{align}
where $E_\gamma$ and $\theta_\gamma$ are the outgoing photon energy and angle, respectively, and $\theta_R$ is the outgoing nuclear recoil angle. The angles are defined to lie between the direction of the incoming neutrino and the outgoing states.

We can again compare the scenarios where the sterile neutrino $N$ is a Dirac or Majorana fermion. For Dirac $N$, an incoming active Dirac neutrino (or Majorana neutrino with negative helicity) triggers the process $\nu_\alpha A\to XA\gamma$. Conversely, an active Dirac antineutrino (or Majorana neutrino with positive helicity) induces the process $\bar{\nu}_\alpha A\to \bar{X}A\gamma$. In each case we neglect the incoming \textquoteleft wrong' helicity Majorana neutrino. The amplitude for the $\nu_\alpha A\to XA\gamma$ process for both Dirac and Majorana $\nu_\alpha$ can thus be written as
\begin{align}\label{eq:Mphoton}
i\mathcal{M}^{\text{D}}_{\nu_\alpha A\to X A \gamma} &= (\mu_{\nu N}^\alpha)^* \mu_{X N}\frac{i\big[\bar{u}_{X}\sigma_{\mu\nu}P_{R}(\slashed{p}\hspace{-0.5em}\phantom{p}_N+m_N)\sigma_{\rho\sigma}P_{L}u_{\nu_\alpha}\big]\epsilon^{\mu*}p_3^{\nu}q^{\sigma}}{p_N^2-m_N^2+i m_N\Gamma_N}\frac{(-ig^{\rho\lambda})}{q^2}\mathcal{J}^{A}_{\lambda}\nonumber\\
&= \mu_{\nu N}^\alpha \mu_{X N}\big[\bar{u}_{X}\sigma_{\mu\nu}P_{R}(\slashed{p}\hspace{-0.5em}\phantom{p}_N+m_N)\sigma_{\rho\sigma}P_{L}u_{\nu_\alpha}\big]F^{\mu\nu\rho\sigma}\,,
\end{align}
where $p_N = p_2 + p_3$, $\Gamma_N$ is the total width of $N$, and $F^{\mu\nu\rho\sigma}\equiv -\frac{\epsilon^{\mu*}p_3^{\nu}(\mathcal{J}^{A})^{\rho}q^{\sigma}}{q^2(p_N^2-m_N^2+i m_N\Gamma_N)}$. For $\bar{\nu}_\alpha A\to\bar{X}A\gamma$ we instead have,
\begin{align}\label{eq:Mphoton2}
i\mathcal{M}^{\text{D}}_{\bar{\nu}_\alpha A\to \bar{X} A \gamma} &= \mu_{\nu N}^\alpha\mu_{X N}\big[\bar{v}_{\nu_\alpha}\sigma_{\rho\sigma}P_{R}(\slashed{p}\hspace{-0.5em}\phantom{p}_N+m_N)\sigma_{\mu\nu}P_{L}v_{X}\big]F^{\mu\nu\rho\sigma}\,.
\end{align}

For Majorana $N$, additional processes are possible. For example, if $\nu_\alpha$ and $X$ are Dirac, the lepton number violating processes $\nu_\alpha A\to \bar{X}A\gamma$ is allowed. The situation is similar if $\nu_\alpha$ and $X$ are Majorana; now the outgoing state $X$ can have negative or positive helicity. However, we emphasise that the outgoing state $X$ is not measured. We then need to sum the matrix elements for the processes $\nu_\alpha A\to XA\gamma$ and $\nu_\alpha A\to \bar{X}A\gamma$ if $X$ is Dirac (or simply the matrix element for $\nu_\alpha A\to XA\gamma$ if $X$ is Majorana), which is
\begin{align}\label{eq:Mphoton3}
\hspace{-0.7em}i\mathcal{M}^{\text{M}}_{\nu_\alpha A\to X A \gamma} &= \mu_{\nu N}^\alpha \mu_{X N}\big[\bar{u}_{X}\sigma_{\mu\nu}(\slashed{p}\hspace{-0.5em}\phantom{p}_N+m_N)\sigma_{\rho\sigma}P_{L}u_{\nu_\alpha}\big]F^{\mu\nu\rho\sigma}\,,
\end{align}
where the decay vertex of $N$ now contains the sum of chirality projectors $(P_R+P_L)=1$. The $P_R$ projector isolates the momentum term $\slashed{p}\hspace{-0.5em}\phantom{p}_N$ in the sterile neutrino propagator, while the $P_L$ projector selects the mass term $m_N$. The former corresponds to the process $\nu_\alpha A\to XA\gamma$ and the latter to $\nu_\alpha A\to \bar{X}A\gamma$. We emphasise that lepton number is not measured in the final state. Considering instead an incoming Dirac $\bar{\nu}_\alpha$ (or a Majorana $\nu_\alpha$ of predominantly positive helicity), the matrix element for is given by Eq.~\eqref{eq:Mphoton3} with the replacement $\big[\bar{u}_{X}\sigma_{\lambda\xi}(\slashed{p}\hspace{-0.5em}\phantom{p}_N+m_N)\sigma_{\mu\rho}P_{L}u_{\nu_\alpha}\big]\to \big[\bar{v}_{\nu_\alpha}\sigma_{\lambda\xi}P_R(\slashed{p}\hspace{-0.5em}\phantom{p}_N+m_N)\sigma_{\mu\rho}v_{X}\big]$.

The differential cross section for the $\nu_\alpha A\to XA\gamma$ process can be found by taking the absolute square of the scattering amplitude, averaging over the possible spins of the incoming nucleus and summing over the spins of the outgoing nucleus. We also sum over the polarisations of the outgoing photon. Neglecting the mass of light neutrinos (and any sterile state $N'$) in the final state, the differential cross section is given by
\begin{align}
\label{eq:sigmaradiative}
d^5\sigma_{\nu_\alpha A\to XA\gamma} &= \frac{1}{2(s-m_A^2)}\frac{1}{2}\sum_{\mathrm{spins}}|\mathcal{M}^{\text{D}(\text{M})}_{\nu_\alpha A\to XA\gamma}|^2 d\Phi_3 \,,
\end{align}
where $d\Phi_3$ is the three-body phase space for the outgoing nucleus, light neutrino and photon,
\begin{align}
\label{eq:3bodyphasespace}
d\Phi_3 &= (2\pi)^4 \frac{d^3 \mathbf{p}_1}{2(2\pi)^3E_{\mathbf{p}_1}}\frac{d^3 \mathbf{p}_2}{2(2\pi)^3E_{\mathbf{p}_2}}\frac{d^3 \mathbf{p}_3}{2(2\pi)^3E_{\mathbf{p}_3}}\delta^{4}(k_1+k_2-p_1-p_2-p_3) \nonumber\\
&= \frac{d\phi}{2\pi}\frac{ds_1 dt_1 ds_3 dt_2}{256\pi^4(s-m_A^2)\sqrt{-\Delta_4}}\,\,.
\end{align}
In the second equality, the integral has firstly been decomposed into a pair of two-body phase spaces (and a trivial integral over the azimuthal orientation of the system) and then written in terms of four Mandelstam variables $s_1$, $t_1$, $s_3$ and $t_2$. The function
\begin{align}
\Delta_4 = -\frac{1}{16}\large\left|\begin{smallmatrix}
0 & 0 & s_3 & t_1 & m_A^2 & 1\\
0 & 0 & 0 & t_2 & s_1 & 1\\
s_3 & 0 & 0 & 0 & s & 1\\
t_1 & t_2 & 0 & 0 & m_A^2 & 1\\
m_A^2 & s_1 & s & m_A^2 & 0 & 1\\
1 & 1 & 1 & 1 & 1 & 0 \\
\end{smallmatrix}\right|\,
\end{align}
is the $4\times 4$ symmetric Gram determinant, with $\Delta_4<0$ defining the physical region of the phase space~\cite{Byckling:1971vca}. 

The spin-averaged and polarisation-summed squared matrix element in the Dirac and Majorana scenarios can be written as
\begin{align}
\label{eq:radiativematrixsq}
\frac{1}{2}\sum_{\text{spins},\,\text{pols}}|\mathcal{M}_{\nu_\alpha A\to X A \gamma}^{\text{D}(\text{M})}|^2 &= |\mu^\alpha_{\nu N}\mu_{X N}|^2 \,\frac{e^2 Z^2\mathcal{F}^2(q^2)}{q^4}\frac{L^{\gamma,\,\text{D}(\text{M})}_{\mu\nu}H^{\mu\nu}}{(p_{N}^2-m_N^2)^2+m_N^2\Gamma_N^2}\,,
\end{align}
where $H^{\mu\nu}$ is given in Eq.~\eqref{eq:Wmunu} and the leptonic parts are
\begin{align}
\label{eq:Lphoton}
L^{\gamma,\,\text{D}}_{\mu\nu} &= \mathrm{Tr}[\slashed{p}\hspace{-0.5em}\phantom{p}_2\sigma_{\lambda\xi}\slashed{p}\hspace{-0.5em}\phantom{p}_N\sigma_{\mu\rho}\slashed{k}\hspace{-0.5em}\phantom{k}_2\sigma_{\nu\omega}P_{R}\slashed{p}\hspace{-0.5em}\phantom{p}_N\sigma_{\eta\zeta}]\,g^{\lambda\eta}p_3^{\xi}p_3^{\zeta}q^{\rho}q^{\omega}\,,\\
L^{\gamma,\,\text{M}}_{\mu\nu} &= \mathrm{Tr}[\slashed{p}\hspace{-0.5em}\phantom{p}_2\sigma_{\lambda\xi}(\slashed{p}\hspace{-0.5em}\phantom{p}_N+m_N)\sigma_{\mu\rho}\slashed{k}\hspace{-0.5em}\phantom{k}_2\sigma_{\nu\omega}P_{R}(\slashed{p}\hspace{-0.5em}\phantom{p}_N+m_N)\sigma_{\eta\zeta}]\,g^{\lambda\eta}p_3^{\xi}p_3^{\zeta}q^{\rho}q^{\omega}\,,
\end{align}
Inserting Eqs.~\eqref{eq:3bodyphasespace} and \eqref{eq:radiativematrixsq} into the cross section formula of Eq.~\eqref{eq:sigmaradiative} now gives the Lorentz-invariant differential cross section
\begin{align}
\label{eq:sigma4}
\frac{d^4\sigma^{\text{D}(\text{M})}_{\nu_\alpha A\to X A \gamma}}{ds_1dt_1ds_3dt_2} &= |\mu^\alpha_{\nu N}\mu_{X N}|^2 \,\frac{\alpha Z^2\mathcal{F}^2(t_1)}{128\pi^3 t_1^2} \,\frac{L^{\gamma,\,\text{D}(\text{M})}_{\mu\nu}H^{\mu\nu}}{(s-m_A^2)^2\left[(s_1 - m_N^2)^2 + 
	m_N^2 \Gamma_N^2\right]\sqrt{-\Delta_4}}\,,
\end{align}
where we have integrated over the azimuthal angle $\phi$. 

\textbf{Connection to observables}:
From Eq.~\eqref{eq:sigma4} we wish to compute the differential cross sections in the experimental observables of interest, in particular the nuclear recoil energy $E_R$, the outgoing photon energy $E_\gamma$ and angle $\theta_\gamma$ (between the incoming neutrino and outgoing photon). Because Eq.~\eqref{eq:sigma4} depends on four Mandelstam variables, we need an other variable in the lab frame, which we choose to be the angle $\theta_R$ (between the incoming neutrino and outgoing recoiling nucleus). We have already given the Mandelstam variables in terms of these labe frame quantities in Eqs.~\eqref{eq:sMandelstam} to \eqref{eq:t2Mandelstam}. We see that $E_R$ and $\theta_R$ only appear in $s_1$ and $t_1$ and $E_\gamma$ and $\theta_\gamma$ in $s_3$ and $t_2$. To determine the single differential cross section in $E_R$, the first step is to therefore integrate Eq.~\eqref{eq:sigma4} over $s_3$ and $t_2$, i.e.
\begin{align}
\label{eq:doublediffs1t1_1}
\frac{d^2\sigma^{\text{D}(\text{M})}_{\nu_\alpha A\to X A \gamma}}{ds_1dt_1} &= \int^{t_{2}^{+}}_{t_{2}^{-}}dt_2 \int^{s_{3}^{+}}_{s_{3}^{-}} ds_3 ~ \frac{d^4\sigma^{\text{D}(\text{M})}_{\nu_\alpha A\to X A \gamma}}{ds_1dt_1ds_3dt_2}\,.
\end{align}
The limits of integration are such that the complete physical region of the phase space, $\Delta_4 <0$, is integrated over. The limits $s^{\pm}_{3}(s_1,t_1,t_2)$ are hence found by solving $\Delta_4 = 0$ for $s_{3}$, while $t^{\pm}_{2}(s_1,t_1)$ are found by solving $s^{+}_{3}=s^{-}_{3}$ for $t_2$. Performing this integration for the full differential cross section in Eq.~\eqref{eq:sigma4}, we obtain for Dirac $N$
\begin{align}
\label{eq:sigmas1t1Dirac}
&\frac{d^2\sigma^{\text{D}}_{\nu_\alpha A\to X A \gamma}}{ds_1dt_1} = |\mu^\alpha_{\nu N}\mu_{X N}|^2\,\frac{\alpha Z^2 \mathcal{F}^2(t_1)}{16\pi^2}\frac{s_1^2 F(t_1,s,m_A^2,s_1)}{t_1^2(s-m_A^2)^2 \left[(s_1 - m_N^2)^2 + m_N^2 \Gamma_N^2\right]} \,,
\end{align}
while for Majorana $N$,
\begin{align}
\label{eq:sigmas1t1Majorana}
&\frac{d^2\sigma^{\text{M}}_{\nu_\alpha A\to X A \gamma}}{ds_1dt_1} = |\mu^\alpha_{\nu N}\mu_{X N}|^2\,\frac{\alpha Z^2 \mathcal{F}^2(t_1)}{16\pi^2}\frac{s_1 (s_1+m_N^2) F(t_1,s,m_A^2,s_1)}{t_1^2(s-m_A^2)^2 \left[(s_1 - m_N^2)^2 + m_N^2 \Gamma_N^2\right]} \,.
\end{align}
Taking the ratio of these cross sections, we see that they vary by the factor
\begin{align}
\bigg(\frac{d^2\sigma^{\text{M}}_{\nu_\alpha A\to X A \gamma}}{ds_1dt_1}\bigg)\bigg/\bigg(\frac{d^2\sigma^{\text{D}}_{\nu_\alpha A\to X A \gamma}}{ds_1dt_1}\bigg) = 1+\frac{m_N^2}{s_1}\,.
\end{align}
The first term corresponds to the process $\nu_\alpha A\to X A \gamma$ which possible for both Dirac and Majorana $N$. The second term instead corresponds to the process $\nu_\alpha A\to \bar{X} A \gamma$ which requires a helicity flip of $N$ and is only possible for Majorana $N$.

\begin{figure}[t!]
	\centering
	\includegraphics[width=0.6\textwidth]{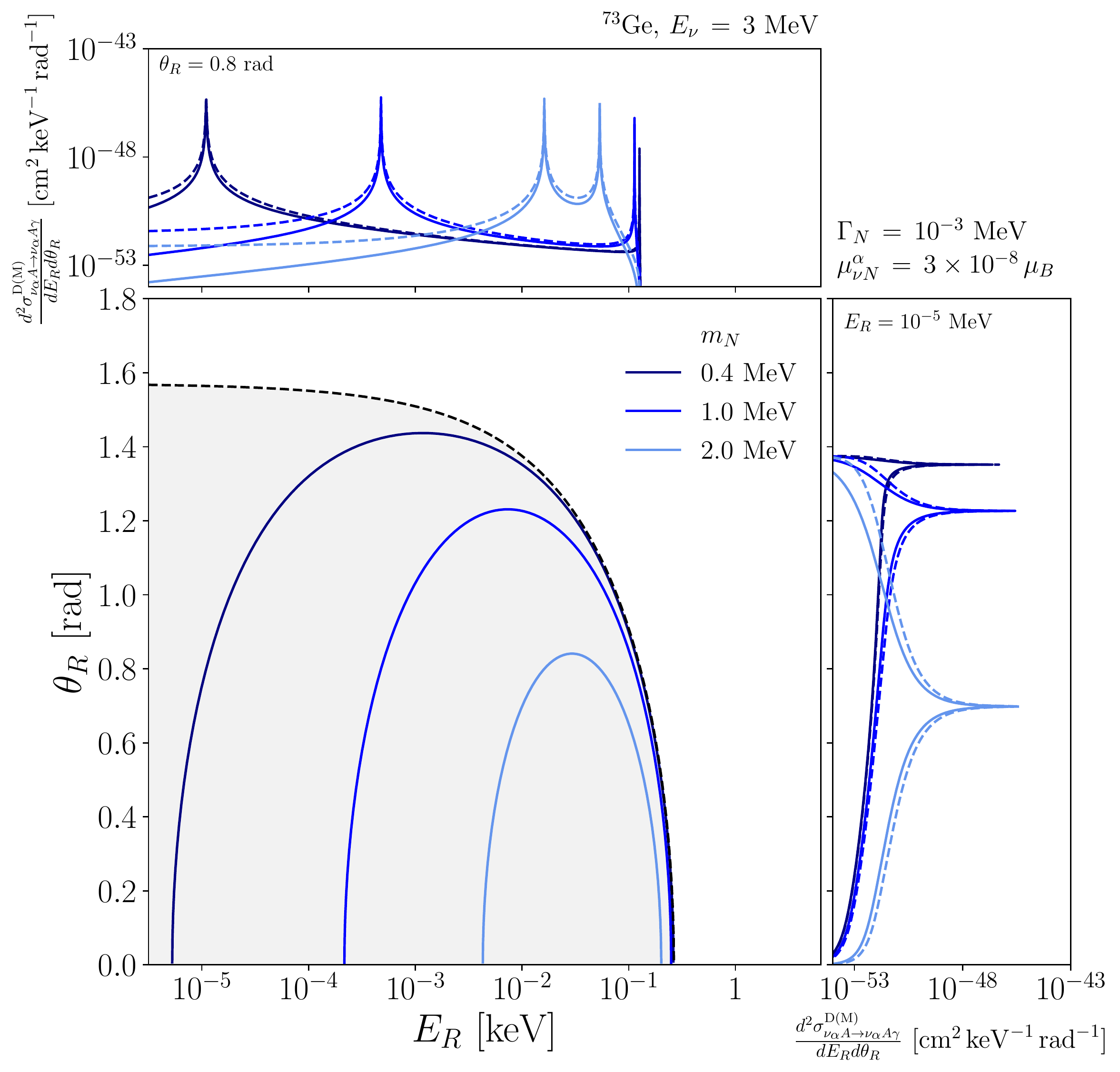}
	\caption{Kinematically allowed region in the $(E_R,\theta_R)$ plane for the $\nu_\alpha A\to \nu_\alpha A \gamma$ process, indicated by the grey shaded region. For three different values of $m_{N}$ the relationship between $E_{R}$ and $\theta_R$ is shown the narrow width approximation (NWA). The side plots depict the double differential cross section in $E_{R}$ and $\theta_{R}$, again for three different values of $m_{N}$ and in the Dirac (solid) and Majorana (dashed) cases. The top plot is for fixed $\theta_{R} = 0.5$~rad and the right for fixed $E_{R}=10^{-5}$~MeV. As $\Gamma_{N} \ll m_{N}$, it can be seen that the differential cross sections are sharply peaked at values of $E_{R}$ and $\theta_{R}$ satisfying the relationship in the NWA.}
	\label{fig:nuclear_param_space}
\end{figure}

To determine the differential cross section in the relevant lab frame quantities $E_R$ and $\theta_R$, we now multiply Eqs.~\eqref{eq:sigmas1t1Dirac} and \eqref{eq:sigmas1t1Majorana} by a Jacobian, i.e.
\begin{align}
\label{eq:doublediffrecoil}
\frac{d^2\sigma^{\text{D}(\text{M})}_{\nu_\alpha A\to X A \gamma}}{dE_Rd\theta_R}&= \left|\frac{\partial(s_1,t_1)}{\partial(E_R,\theta_R)}\right|\,\frac{d^2\sigma^{\text{D}(\text{M})}_{\nu_\alpha A\to X A \gamma}}{ds_1dt_1}\,,
\end{align}
where $\big|\frac{\partial(s_1,t_1)}{\partial(E_R,\theta_R)}\big| = -4 E_\nu m_A \sqrt{E_R (2 m_A + E_R)} \sin\theta_R$.
The differential cross sections in the nuclear recoil energy $E_R$ and angle $\theta_R$ can now be computed by integrating over the remaining variable,
\begin{align}
\frac{d\sigma^{\text{D}(\text{M})}_{\nu_\alpha A\to X A \gamma}}{dE_R} &= \int^{\theta^{+}_{R}}_{0} d\theta_R ~  \frac{d^2\sigma^{\text{D}(\text{M})}_{\nu_\alpha A\to X A \gamma}}{dE_Rd\theta_R}\,,\\
\frac{d\sigma^{\text{D}(\text{M})}_{\nu_\alpha A\to X A \gamma}}{d\theta_R} &= \int^{E^{+}_{R}}_{0} dE_R ~  \frac{d^2\sigma^{\text{D}(\text{M})}_{\nu_\alpha A\to X A \gamma}}{dE_Rd\theta_R}\,.
\end{align}
The allowed region is bounded by the upper limits 
\begin{align}
\cos\theta_{R}^{+} = \frac{E_R(m_A+E_\nu)}{E_\nu\sqrt{E_R(2m_A+E_R)}}\,, \quad E_{R}^{+} = \frac{2 m_A E_\nu^2 \cos^2\theta_R}{m_A(m_A+2 E_\nu)+E_\nu^2(1-\cos^2\theta_R)}\,.
\end{align}
In Fig.~\ref{fig:nuclear_param_space} we depict the kinematically-allowed region as the grey shaded area in the main plot. The kinematically allowed region can be seen to be independent of the sterile neutrino mass $m_N$. In the sub-plots above and to the right, we show the double differential cross section in Eq.~\eqref{eq:doublediffrecoil} for fixed $\theta_R = 0.5$~rad (above) and $E_R = 10^{-5}$~MeV (right) and three different values of $m_N$. For illustrative purposes, we set the values of the transition magnetic moment and total decay width of $N$ to be $\mu^{\alpha}_{\nu N} = 3\times 10^{-8}$~$\mu_B$ and $\Gamma_N = 10^{-3}$~MeV, respectively. A total decay width of this size would require additional invisible decay modes of $N$. We observe that, even though the double differential cross sections are non-zero over the entire kinematically allowed region, they are dominated by sharply peaked regions. This is a consequence of the total decay width of $N$ being much smaller than the mass of $N$, justifying the use of the narrow width approximation (NWA). 

In the $\Gamma_N \ll m_N$ limit, the following replacement can be made in Eq.~\eqref{eq:sigma4},
\begin{align}
\label{eq:NWA_sub}
\frac{1}{(s_1 - m_N^2)^2 + m_N^2 \Gamma_N^2} \to \frac{\pi}{m_N\Gamma_N}\delta(s_1-m_N^2)
\end{align}
which sets the intermediate $N$ to be on-shell, i.e. $s_1 = p_N^2 = m_N^2$. Inserting the expression of $s_1$ in terms of the lab frame variables in Eq.~\eqref{eq:s1Mandelstam} into $s_1 = m_N^2$ allows to find the following relationship between the nuclear recoil angle and energy,
\begin{align}
\cos\theta_R \big|_{\text{NWA}}= \frac{m_N^2 + 2 E_R (m_A + E_\nu)}{2 E_\nu \sqrt{E_R (2 m_A + E_R)}}\,,
\end{align}
or equivalently,
\begin{align}
\label{eq:ERpm}
E_{R}^{\pm}\big|_{\text{NWA}} &= \frac{2 m_A E_\nu^2  c^2_{R}-m_N^2(m_A + E_\nu) \pm E_\nu c_{R}\sqrt{4 m_A^2 E_\nu^2  c^2_{R} - 4 m_A m_N^2 (m_A + E_\nu) + m_N^4}}{2 m_A (m_A + 2E_\nu)+2E_\nu^2 (1-c^2_{R})}\,,
\end{align}
where $c_R = \cos\theta_R$. In Fig.~\ref{fig:nuclear_param_space} we plot the curves of allowed values in the $(E_R, \theta_R)$ plane from the condition $s_1 = m_N^2$ for three different values of $m_N$. From the plots above and to the right, we see that the double differential cross section is sharply peaked at these values of $E_R$ and $\theta_R$. For each value of $m_{N}$ there is an maximum recoil angle,
\begin{align}
\cos\theta_{R}^{\mathrm{max}}\big|_{\text{NWA}} = \frac{m_N^2\big(m_A + E_\nu + \sqrt{m_A (m_A + 2 E_\nu)}\big)}{2 E_\nu \big(m_A m_N^2 \sqrt{m_A (m_A + 2 E_\nu)} + m_N^4\big)^{1/2}}\,,
\end{align}
as well as minimum and maximum recoil energies given by Eq.~\eqref{eq:ERpm} with \mbox{$c_R = \pm 1$}. We now take Eqs.~\eqref{eq:sigmas1t1Dirac} and \eqref{eq:sigmas1t1Majorana}, make the substitution in Eq.~\eqref{eq:NWA_sub} and integrate over $s_1$ by setting $s_1 = m_N^2$. To obtain the differential cross section in the nuclear recoil energy, we finally multiply by the Jacobian factor $\partial t/\partial E_R = -2 m_A$ to obtain
\begin{align}
\label{eq:radiative_cross_NWA}
\frac{d\sigma^{\text{D}(\text{M})}_{\nu_\alpha A\to X A \gamma}}{dE_R}\bigg|_{\text{NWA}} = \frac{d\sigma_{\nu_\alpha A\to N A}}{dE_R}\frac{\Gamma^{\text{D}(\text{M})}_{N\to X\gamma}}{\Gamma_{N}}\,.
\end{align}
In the NWA, the differential rate in the nuclear recoil for the $\nu_\alpha A\to X A \gamma$ process is therefore the Primakoff upscattering cross section multiplied by the branching ratio for the radiative decay $N\to X\gamma$.

To obtain the differential cross sections for $\nu_\alpha A\to X A \gamma$ in the outgoing photon energy $E_\gamma$ and angle $\theta_\gamma$, we can instead integrate Eq.~\eqref{eq:sigma4} over $s_1$ and $t_1$ as
\begin{align}
\label{eq:integrate_t1s1}
\frac{d^2\sigma^{\text{D}(\text{M})}_{\nu_\alpha A\to X A \gamma}}{ds_3dt_2} &= \int^{t^{+}_1}_{t^{-}_1} dt_1\int^{s^{+}_1}_{s^{-}_1} ds_1 ~ \frac{d^4\sigma^{\text{D}(\text{M})}_{\nu_\alpha A\to X A \gamma}}{ds_1dt_1ds_3dt_2}\,,
\end{align}
where again the limits $s^{\pm}_1(t_1,s_3,t_2)$ are found by solving $\Delta_4 = 0$ for $s_1$ and $t^{\pm}_1(s_3,t_2)$ by solving $s^{+}_1 = s^{-}_1$ for $t_1$. In general both integrals are non-trivial for the differential cross section in Eq.~\eqref{eq:sigma4}  because $s_1$ appears in the factor of $[(s_1-m_N^2)+m_N^2\Gamma_N^2]$ in the denominator and $t_1$ appears in the nuclear form-factor $\mathcal{F}(t_1)$; it is therefore necessary to perform this integration numerically. Once this is done, it is possible to transform to the lab frame by multiplying Eq.~\eqref{eq:integrate_t1s1} by the Jacobian $\big|\frac{\partial(s_3,t_2)}{\partial(E_\gamma,\theta_\gamma)}\big| = 4 m_A E_\nu E_\gamma \sin\theta_\gamma$.
To obtain the single differential cross sections in the variables $E_\gamma$ and $\theta_\gamma$, the remaining variables are integrated over as
\begin{align}
\label{eq:diff_Egamma}
\frac{d\sigma^{\text{D}(\text{M})}_{\nu_\alpha A\to X A \gamma}}{dE_\gamma}&=  \int^{\pi}_{0} d\theta_{\gamma}~\frac{d^2\sigma^{\text{D}(\text{M})}_{\nu_\alpha A\to X A \gamma}}{dE_\gamma d\theta_\gamma}\,,\\
\frac{d\sigma^{\text{D}(\text{M})}_{\nu_\alpha A\to X A \gamma}}{d\theta_\gamma} &=  \int^{E^{+}_{\gamma}}_{0} dE_{\gamma}~\frac{d^2\sigma^{\text{D}(\text{M})}_{\nu_\alpha A\to X A \gamma}}{dE_\gamma d\theta_\gamma}\,,
\label{eq:diff_thetagamma}
\end{align}
where the maximum allowed photon energy is $E^{+}_\gamma = \frac{m_A E_\nu}{m_A + E_\nu(1-\cos\theta_\gamma)}$.
\begin{figure*}[t!]
	\centering
	\includegraphics[width=0.46\textwidth]{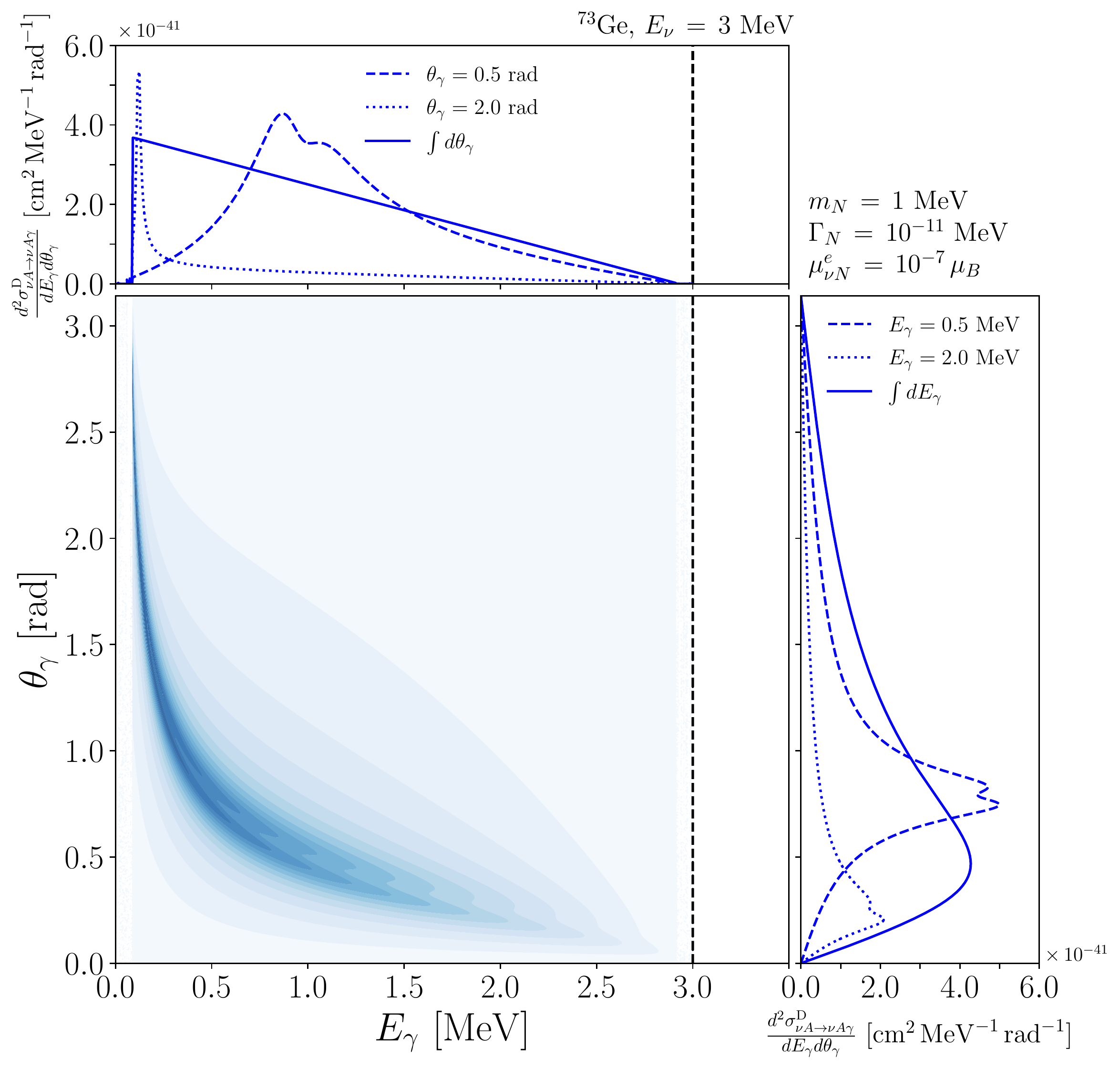}
	\includegraphics[width=0.46\textwidth]{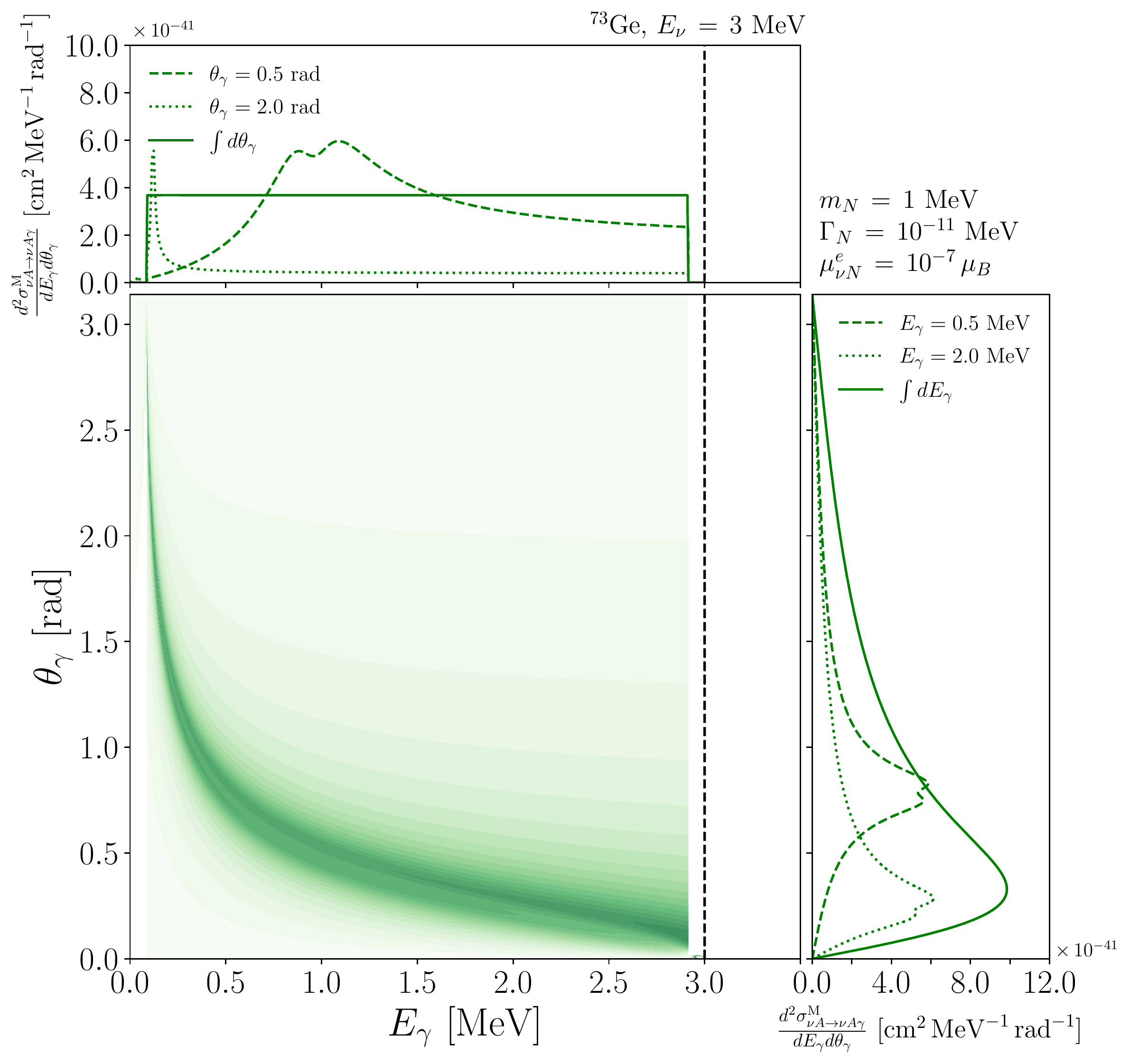}
	\caption{Kinematically allowed region in the $(E_\gamma,\theta_\gamma)$ plane for the $\bar{\nu}_e A\to \bar{\nu}_e A \gamma$ process, indicated by the blue (Dirac case, left) and green (Majorana case, right) shaded regions to the left of the black dashed line. The incoming neutrino energy, nuclear target and sterile neutrino parameters are indicated in the plots. As in Fig.~\ref{fig:contour_plots}, the contours depict the size of the double differential cross section in $E_\gamma$ and $\theta_\gamma$. The side plots now depict the double differential cross sections for fixed values of $E_\gamma$ (right) and $\theta_\gamma$ (above) as dashed and dotted lines. Also shown are the single differential cross sections found by integrating over all allowed values of the other variable (solid lines).}
	\label{fig:photon_param_space}
\end{figure*}

However, the NWA can also be used to simplify the calculation above. The substitution in Eq.~\eqref{eq:NWA_sub} can be made in Eq.~\eqref{eq:integrate_t1s1} and the $s_1$ integration performed by setting $s_1 = m_N^2$. However, the $t_1$ integral must be still be performed numerically due to the non-trivial dependence of $\mathcal{F}(t_1)$.  Multiplying the double differential cross section by the Jacobian $\big|\frac{\partial(s_3,t_2)}{\partial(E_\gamma,\theta_\gamma)}\big|$, we obtain the double differential cross section in $E_\gamma$ and $\theta_\gamma$,
\begin{align}
\hspace{-0.5em}\frac{d^2\sigma^{\text{D}(\text{M})}_{\nu_\alpha A\to X A \gamma}}{dE_\gamma d\theta_\gamma}\bigg|_{\text{NWA}} &= \frac{|\mu^\alpha_{\nu N}\mu_{X N}|^2\alpha Z^2 E_\gamma\sin\theta_\gamma}{128\pi^2 m_A E_\nu m_N \Gamma_N }\int^{t^{+}_1}_{t^{-}_1} dt_1 \, \frac{L_{\mu\nu}^{\gamma,\,\text{D(M)}}H^{\mu\nu} \mathcal{F}^2(t_1)}{t_1^2 \sqrt{-\Delta_4}}\bigg|_{s_1 = m_N^2}\,.
\end{align}
In the following, we set $\mathcal{F}(t_1) = 1$ and perform the integral over $t_1$ analytically. In the NWA, the limits of integration $t_1^{\pm}$ can be found by solving $\Delta_4 = 0$ for $t_1$ with $s_1 = m_N^2$. These limits correspond to the minimum and maximum photon energies
\begin{align}
\label{eq:E_gamma_minmax}
E^{-}_\gamma\big|_{\text{NWA}} &= \frac{2 m_A E_\nu + m_N^2 - \sqrt{4 E_{\nu}^2 m_A^2 - 4 m_A (m_A + E_{\nu}) m_N^2 + m_N^4}}{4(m_A + 2E_\nu)}\,, \\
E^{+}_\gamma\big|_{\text{NWA}} &= \frac{2 m_A E_\nu + m_N^2 + \sqrt{4 E_{\nu}^2 m_A^2 - 4 m_A (m_A + E_{\nu}) m_N^2 + m_N^4}}{4m_A}\,.
\end{align}
For $E_\nu, m_N \ll m_A$, these give the simple result $E_\gamma^{\pm}\big|_{\text{NWA}} \approx \frac{E\nu}{2}\big(1\pm\sqrt{1-\frac{m_N^2}{E_\nu^2}}\big)$.

In Fig.~\ref{fig:photon_param_space}, we plot the double differential cross sections in $E_\gamma$ and $\theta_\gamma$ for Dirac (left) and Majorana (right) $N$. We choose the values $m_N = 1$~MeV, $\Gamma_N = 10^{-11}$~MeV and $\mu_{\nu N}^\alpha = 10^{-7}$~$\mu_B$. In the sub-plots above and to the right of the contours, we also plot the double differential cross section for fixed values of the photon energy $E_\gamma$ (right) and angle $\theta_\gamma$ (above). Furthermore, we plot the single differential cross sections in $E_\gamma$ and $\theta_\gamma$ by integrating over the other variable as in Eqs.~\eqref{eq:diff_Egamma} and \eqref{eq:diff_thetagamma}. Examining the single differential cross sections in the photon energy $E_\gamma$, we see a stark difference between the Dirac and Majorana cases. In the former case, the cross section decreases linearly with the energy, while in the latter the cross section is constant. The minimum and maximum photon energies in Eq.~\eqref{eq:E_gamma_minmax} can clearly be seen. Looking at the single differential cross sections in the photon angle $\theta_\gamma$, the difference between the Dirac and Majorana cases is less prominent; the Majorana cross section peaks at slightly lower angles compared to the Dirac cross section.

\textbf{Differential rates}: With the differential cross section for the Primakoff upscattering in Eq.~\eqref{eq:sigmalabframe}, we can now calculate the differential rate of nuclear recoil events in a CE$\nu$NS experiment as
\begin{align}
\label{eq:Primakoff_rate}
\frac{dR_{\nu_\alpha A\to NA}}{dE_R} = \frac{1}{A\cdot m_p}\int^{E_\nu^{\text{max}}}_{E_\nu^{\text{min}}(E_R)} dE_\nu\, \frac{d\phi_{\nu_\alpha}}{dE_\nu} \,\frac{d\sigma_{\nu_\alpha A\to NA}}{dE_R}\,,
\end{align}
where $\frac{d\phi_{\nu_\alpha}}{dE_\nu}$ is the flux of incoming neutrinos $\nu_\alpha$ per cm$^2$ per second and $E_\nu^{\text{min}}$ is the minimum incoming neutrino energy that can produce a sterile neutrino of mass $m_N$ and a nuclear recoil energy $E_R$,
\begin{align}
\label{eq:Enumin}
E^{\text{min}}_\nu(E_R) = \bigg(\frac{E_R}{2}+\frac{m_N^2}{4m_A}\bigg)\bigg(1+\sqrt{1+\frac{2m_A}{E_R}}\bigg)\,.
\end{align}
In Eq.~\eqref{eq:Primakoff_rate} we have divided by the mass number $A$ of the target isotope multiplied by the proton mass $m_p$ to determine the differential rate per unit mass of the target material. 

Similarly, the differential rate for nuclear recoil events that are \textit{coincident} with an outgoing photon in the detector is given (in the NWA) by
\begin{align}
\label{eq:radiative_rate}
\frac{dR^{\text{D}(\text{M})}_{\nu_\alpha A\to X A\gamma}}{dE_R} = \frac{1}{A\cdot m_p}\int^{E_\nu^{\text{max}}}_{E_\nu^{\text{min}}(E_R)} dE_\nu\, \frac{d\phi_{\nu_\alpha}}{dE_\nu} \,\frac{d\sigma_{\nu_\alpha A\to N A}}{dE_R}\frac{\Gamma^{\text{D}(\text{M})}_{N\to X\gamma}}{\Gamma_{N}}P_{N}^{\text{det}}\,.
\end{align}
Here, we simply integrate over the flux $\frac{d\phi_{\nu_\alpha}}{dE_\nu}$ multiplied by the radiative upscattering cross section and the probability $P_{N}^{\text{det}}=1-\text{exp}(-\frac{L_{\text{det}} \Gamma_{N}}{\beta\gamma})$ for the decay to take place inside the detector. In the NWA, $E^{\text{min}}_\nu(E_R)$ is again given by Eq.~\eqref{eq:Enumin}. 

\begin{figure}[t!]
	\centering
	\includegraphics[width=0.42\textwidth]{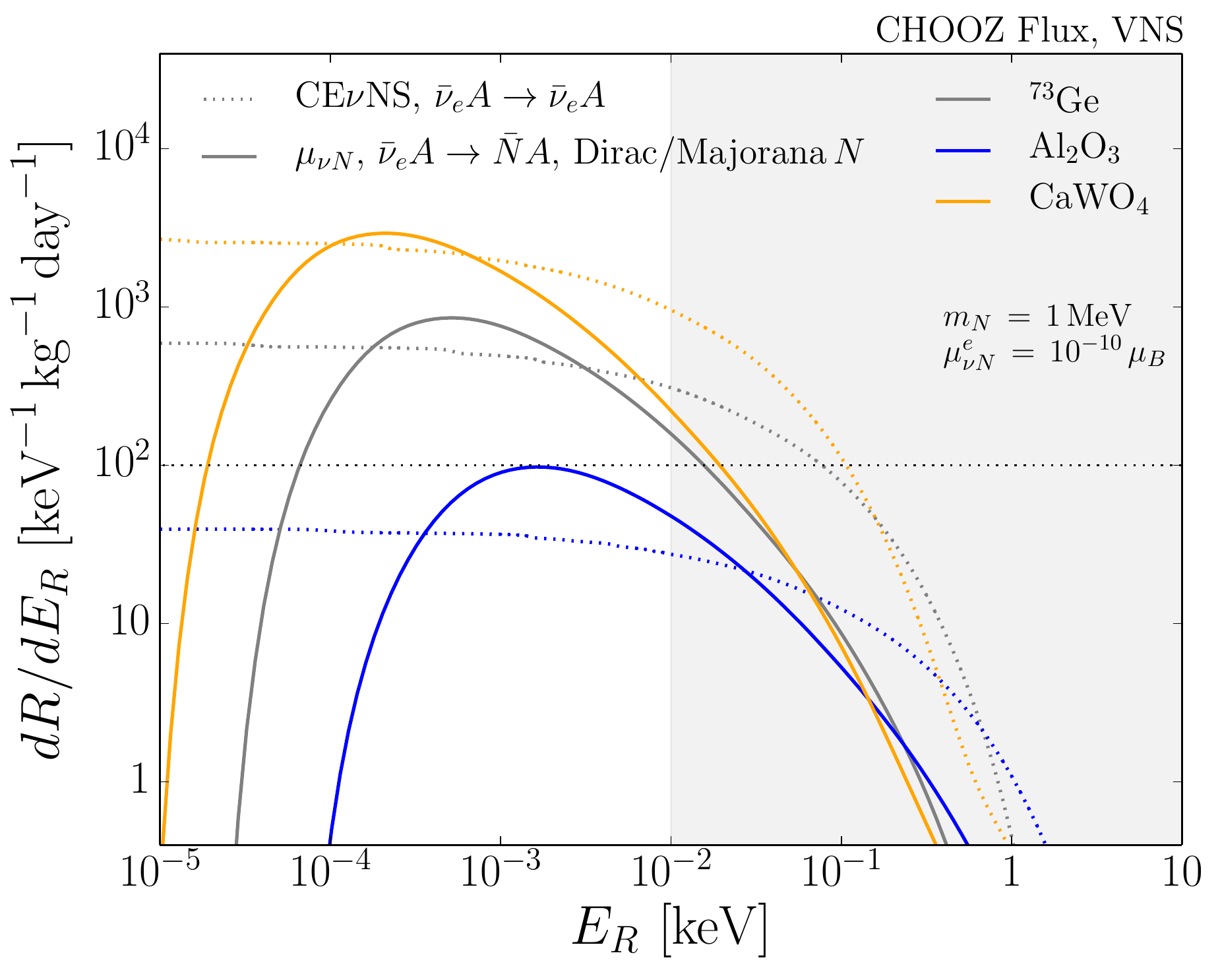}
	\includegraphics[width=0.42\textwidth]{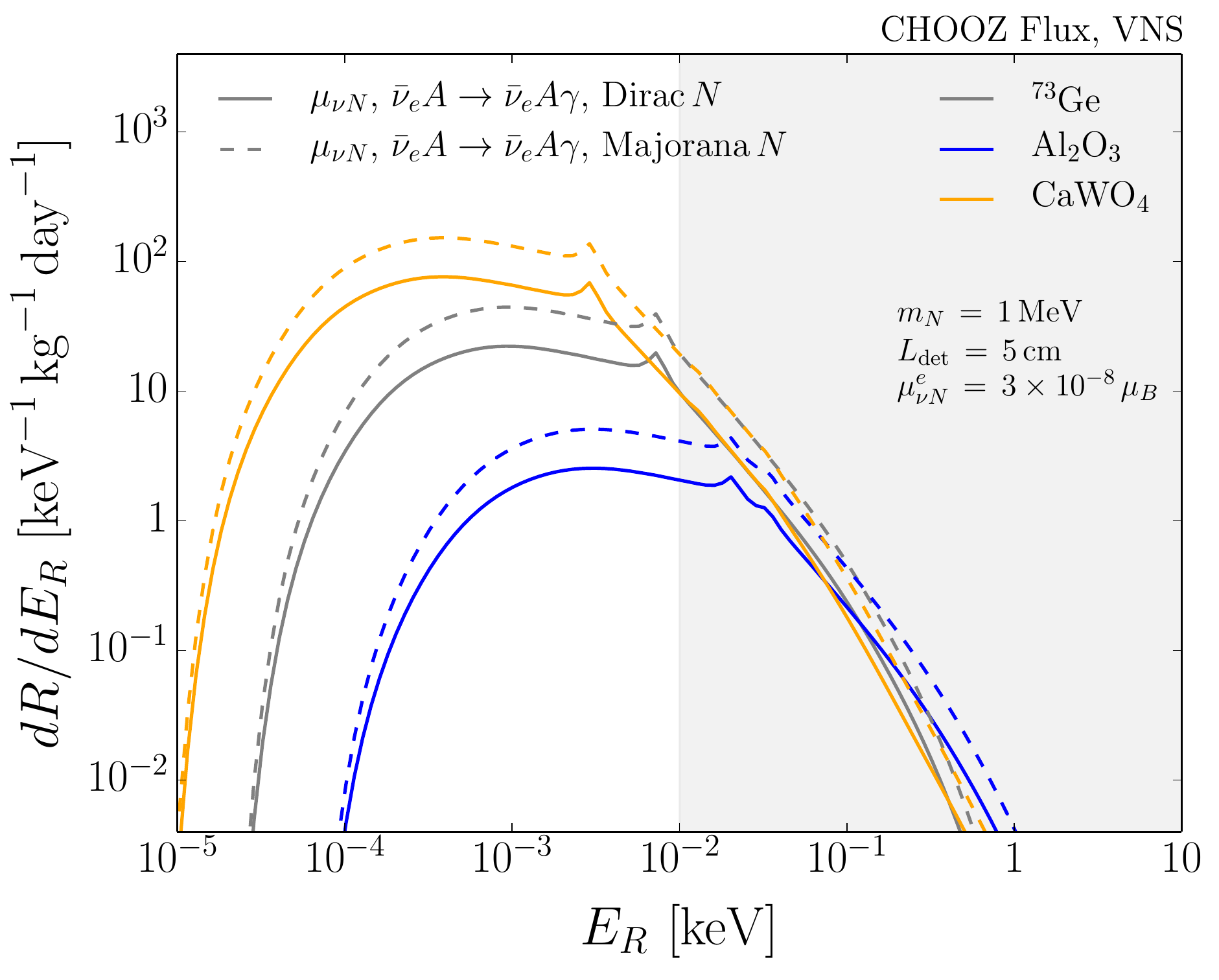}
	\caption{(Left) Differential rates in the nuclear recoil energy $E_R$ for $\bar{\nu}_e A\to \bar{N} A$ per kilogram of a detector at the VNS of the Chooz reactor site, for $m_{N} = 1\,\,\mathrm{MeV}$ and $\mu^e_{\nu N} = 10^{-10}\,\,\mu_B$. These are compared to the SM CE$\nu$NS rates (dotted lines). (Right) Differential rates in $E_R$ for $\bar{\nu}_e A\to \bar{\nu}_e A \gamma$  per kilogram of detector, for $m_{N} = 1\,\,\mathrm{MeV}$ and  $\mu^e_{\nu N} = 3\times 10^{-8}\,\,\mu_B$. The sterile neutrino, which may be a Dirac (solid lines) or Majorana (dashed) fermion, is required to decay inside the detector with $L_{\text{det}}=5 \,\,\mathrm{cm}$. Three possible detector materials are shown; $^{73}$Ge (grey), $\mathrm{Al}_2 \mathrm{O}_3$ (blue) and $\mathrm{CaWO}_4$ (orange).}
	\label{fig:nuclear_recoil_plots}
\end{figure}

As an example, in Fig.~\ref{fig:nuclear_recoil_plots} (left) we plot the differential Primakoff upscattering rate in the nuclear recoil energy for the NUCLEUS experiment situated at the very-near-site (VNS) of the Chooz reactor site, for $m_N = 1$~MeV and $\mu^e_{\nu N}=10^{-10}$~$\mu_B$. The flux of electron antineutrinos induces the process $\bar{\nu}_e A \to \bar{N} A$. In the plot we compare the rates for three different target materials; $^{73}$Ge (grey), $\mathrm{Al}_2 \mathrm{O}_3$ (blue) and $\mathrm{CaWO}_4$ (orange). For $\mathrm{Al}_2 \mathrm{O}_3$ and $\mathrm{CaWO}_4$ we average over the mass numbers of the constituent isotopes. We also compare the Primakoff upscattering rates to the SM CE$\nu$NS process $\bar{\nu}_e A \to \bar{\nu}_e A$ (dotted lines). For $\mu^e_{\nu N}=10^{-10}$~$\mu_B$, the number of Primakoff upscattering events is comparible the number of CE$\nu$NS events. The horizontal black dotted line indicates the predicted nuclear recoil background of 100~keV$^{-1}$~kg$^{-1}$~day$^{-1}$ in the NUCLEUS detector at the VNS. The grey shaded region shows the range of nuclear recoils that can be detected by the experiment (i.e. a nuclear recoil threshold of 10 eV). 

In Fig.~\ref{fig:nuclear_recoil_plots} (right) we plot the differential rate in the nuclear recoil energy for the radiative upscattering process $\bar{\nu}_e A \to \bar{\nu}_e A \gamma$, again for the NUCLEUS experiment at the VNS and for $m_N = 1$~MeV and $\mu^e_{\nu N}=3\times10^{-8}$~$\mu_B$. We assume that the outgoing antineutrino is of electron flavour, $\bar{\nu}_e A \to \bar{\nu}_e A \gamma$ (i.e. there are no additional decay modes of $N$), and that the nuclear recoil is accompanied by an outgoing photon. We compare the cases where the intermediate sterile neutrino $N$ is a Dirac (solid lines) or Majorana (dashed lines) fermion, again for three different target materials. 

One may first notice that there are peaks in these distributions, which otherwise have the same shape as the Primakoff upscattering distributions. The origin of these peaks is as follows; as we assume that $N$ can only decay radiatively, the branching ratio in the integrand of Eq.~\eqref{eq:radiative_rate} is unity. For $\mu^e_{\nu N}=3\times10^{-8}$~$\mu_B$ and $L_{\text{det}} = 5$~cm, the total width satisfies $\Gamma_N = \Gamma^{\text{D}(\text{M})}_{N\to\bar{\nu}_e\gamma}\ll \beta\gamma/L_{\text{det}}$ and the decay probability can be Taylor expanded to give $P_{N}^{\text{det}} \approx L_{\text{det}}\Gamma_N/\beta\gamma$. The peaks occur at values of the nuclear recoil energy that minimise the minimum incoming neutrino energy in Eq.~\eqref{eq:Enumin} and hence minimise $\beta\gamma = \sqrt{\gamma^2-1} \approx \sqrt{\big(\frac{E_\nu}{m_N}\big)^2-1}$. This in turn maximises $P_{N}^{\text{det}}$ and the integrand in Eq.~\eqref{eq:radiative_rate}. In other words, for these values of the nuclear recoil, the incoming neutrino energy is as close as possible to $m_N$. The produced sterile state is non-relativistic and has a shorter decay length $\ell_N = \beta\gamma \tau_N = \beta\gamma/\Gamma_N$ than average. If more decays occur inside the detector, we would expect to observe a peak in coincidence events.

\begin{figure}[t!]
	\centering
	\includegraphics[width=0.42\textwidth]{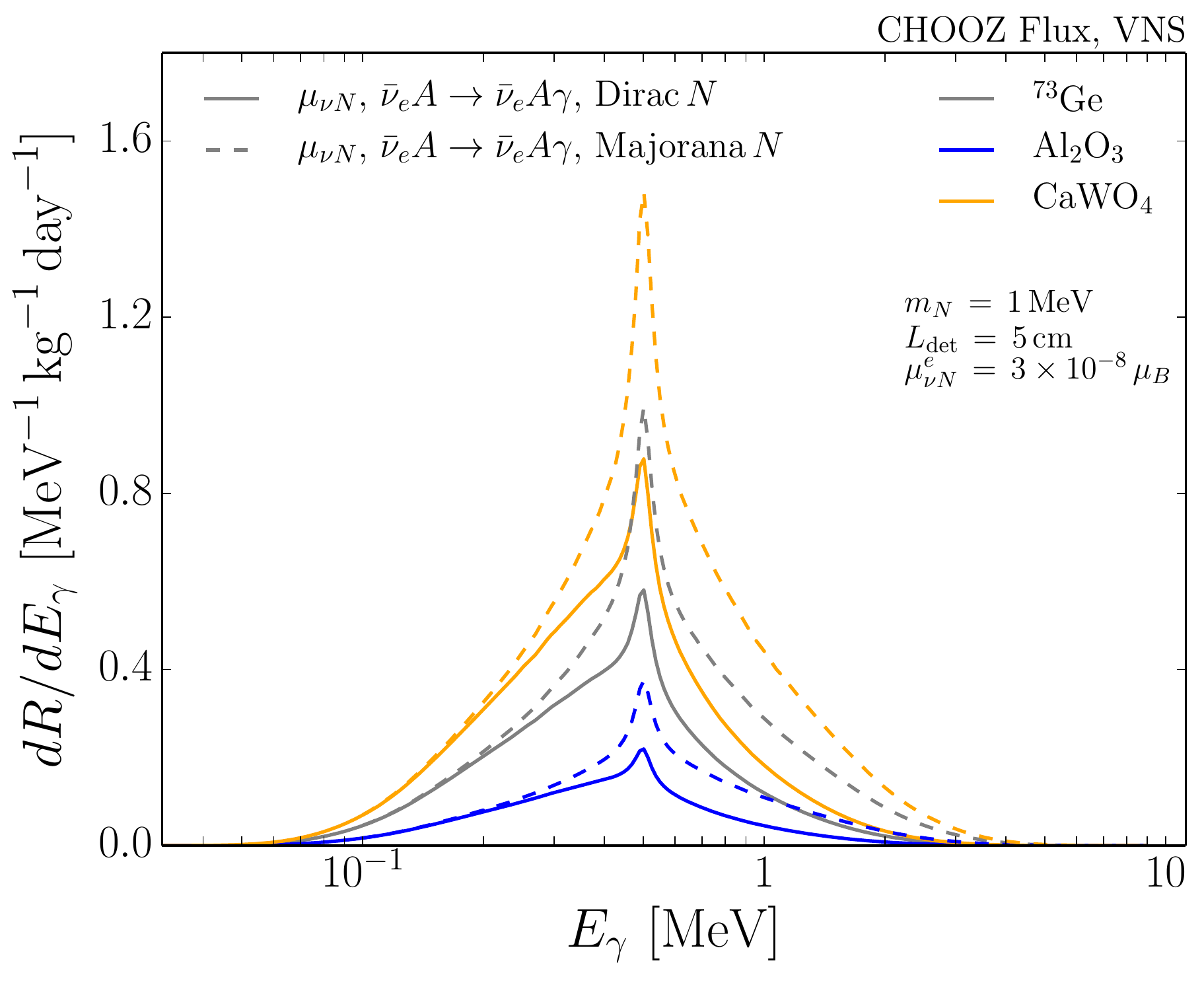}
	\includegraphics[width=0.43\textwidth]{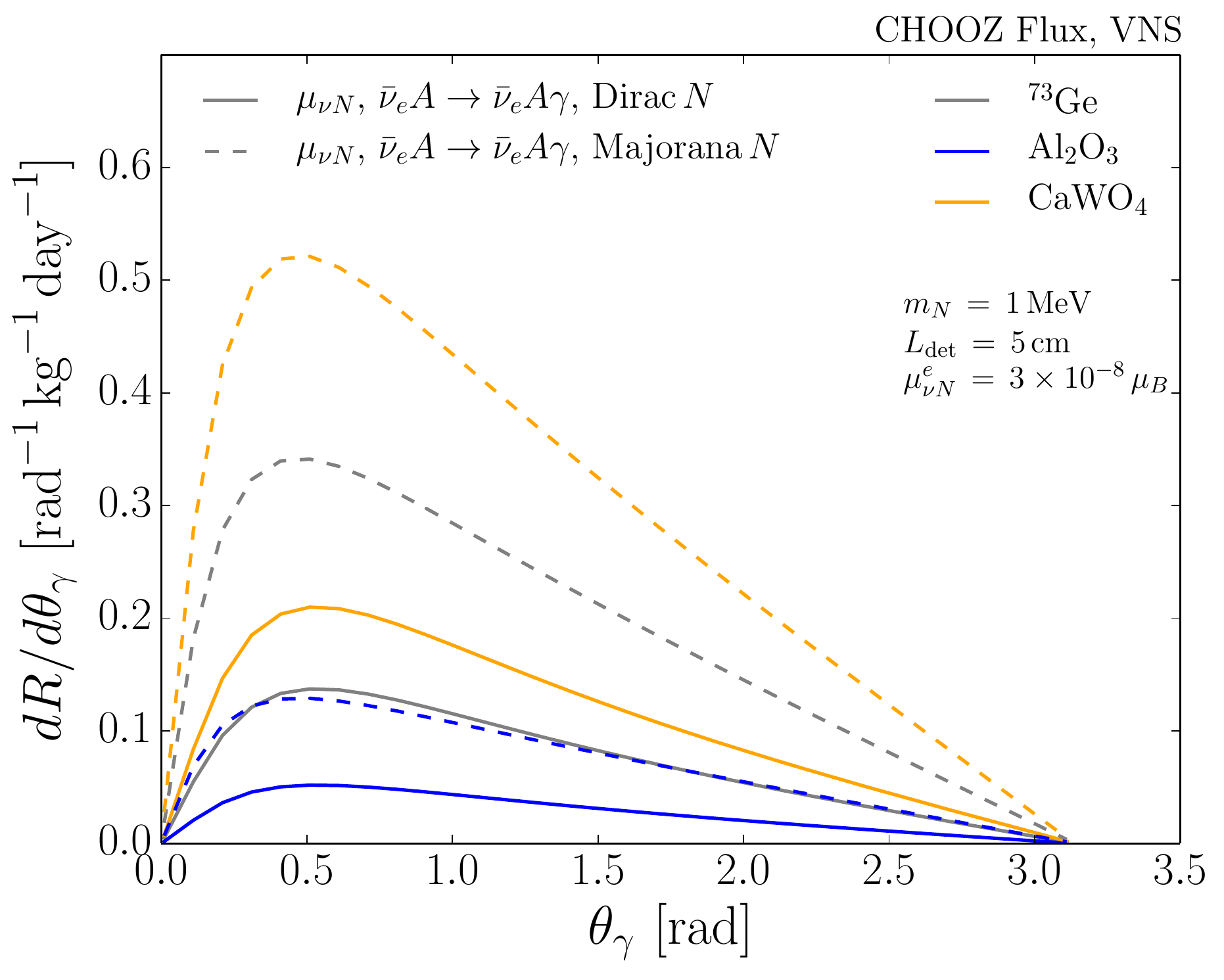}
	\caption{Differential rates of coincidence events in the photon energy $E_{\gamma}$ (left) and photon angle $\theta_{\gamma}$ (right) per kg of a detector situated at the VNS of the Chooz reactor site. Three possible detector materials are shown; $^{73}$Ge (grey), $\mathrm{Al}_2 \mathrm{O}_3$ (blue) and $\mathrm{CaWO}_4$ (orange). The differential rates are different in the Dirac (solid) and Majorana (dashed) cases. The sterile neutrino mass and transition magnetic moment are chosen to be $m_{N} = 1\,\,\mathrm{MeV}$ and  $\mu^e_{\nu N} = 3\times 10^{-8}\,\,\mu_B$, respectively. For a coincidence event to be observed, the sterile neutrino is required to decay inside the detector with $L_{\text{det}}=5 \,\,\mathrm{cm}$.}
	\label{fig:CHOOZ_photon_plots}
\end{figure}

It is also useful to obtain the differential rate for the radiative upscattering process in $E_\gamma$ or $\theta_\gamma$ as
\begin{align}
\frac{dR^{\text{D}(\text{M})}_{\nu_\alpha A\to X A\gamma}}{dX_\gamma} = \frac{1}{A\cdot m_p}\int^{E_\nu^{\text{max}}}_{E_\nu^{\text{min}}(X_\gamma)} dE_\nu\, \frac{d\phi_{\nu_\alpha}}{dE_\nu} \,\frac{d\sigma^{\text{D}(\text{M})}_{\nu_\alpha A\to X A \gamma}}{dX_\gamma}P_{N}^{\text{det}}\,,
\end{align}
where $X_\gamma = \{E_\gamma,\theta_\gamma\}$. In the NWA, the minimum incoming neutrino energy that can produce sterile neutrino of mass $m_N$ and a photon of energy $E_\gamma$ is
\begin{align}
E^{\text{min}}_\nu(E_\gamma)\big|_{\text{NWA}} = E_\gamma + \frac{m_A m_N^2}{4 m_A E_\gamma - 2m_N^2}\,.
\end{align}
For a given incoming neutrino energy $E_\nu$ and sterile neutrino mass $m_N$, the outgoing photon can be emitted at any angle in the range $\theta_\gamma \in [0,\pi]$. In the NWA, the minimum incoming neutrino energy is that which can produce a sterile neutrino of mass $m_N$,
\begin{align}
E^{\text{min}}_\nu(\theta_\gamma)\big|_{\text{NWA}} = \frac{m_N(2m_A - m_N)}{2 (m_A-m_N)}\,.
\end{align}

In Fig.~\ref{fig:CHOOZ_photon_plots} we show the differential rates for the $\bar{\nu}_e A\to \bar{\nu}_e A\gamma$ process in the outgoing photon energy $E_{\gamma}$ (left) and photon angle $\theta_{\gamma}$ (right) for a detector situated at the VNS of the Chooz reactor site. We again present the distributions for three target materials: $^{73}$Ge (grey), $\mathrm{Al}_2 \mathrm{O}_3$ (blue) and $\mathrm{CaWO}_4$ (orange). For the Chooz reactor neutrino flux, the maximum number of events are expected for sterile neutrino masses $m_N\sim 1\-- 5$~MeV. We want to emphasise that the Dirac and Majorana cases (solid and dashed lines, respectively) have different distributions in the photon energy and angle. We again observe enhancements in Fig.~\ref{fig:CHOOZ_photon_plots} (left) at values of $E_\gamma$ that minimise the minimum incoming neutrino energy, i.e. $E^{\text{min}}_\nu(E_\gamma) \sim m_N$.

\bibliography{references}
\end{document}